\documentclass[12pt,twoside,a4paper]{article}
\usepackage{amsmath,amssymb,latexsym,theorem,natbib,epsfig,color}
\usepackage{multirow,graphicx,array,rotating}  
\usepackage{url,afterpage}
\usepackage[normalem]{ulem}
\newfont{\msa}{msam10 scaled\magstep1}
\newfont{\ssmsa}{msam9}

\def\crps{\mathop{\hbox{\rm CRPS}}}

\def\twcrps{\mathop{\hbox{\rm twCRPS}}}

\def\bs{\mathop{\hbox {\rm BS}}}

\def\md{\mathrm{MD}}

\def\es{\mathop{\hbox{\rm ES}}}

\def\permille{\ensuremath{{}^\text{o}\mkern-5mu/\mkern-3mu_\text{oo}}}

\numberwithin{equation}{section}

\title{Statistical post-processing of heat index ensemble forecasts: 
is there a royal road?}

\author{{\sc S\'andor Baran$^{1,2}$},  {\sc \'Agnes Baran$^{2}$}, {\sc Florian Pappenberger$^{1}$} \\ and {\sc Zied Ben Bouall\`egue$^{1}$} \vspace*{0.5cm}\\
$^1$European Centre for Medium-Range Weather Forecasts\\ 
         Shinfield Park, Reading, RG2 9AX, United Kingdom \\
         $^2$Faculty of Informatics, University of Debrecen\\
         Kassai \'ut 26, H-4028 Debrecen, Hungary 
        }

        \date{}

\begin{document}
\pagestyle{myheadings}

\maketitle

\begin{abstract}
We investigate the effect of statistical post-processing on the probabilistic skill of discomfort index (DI) and indoor wet-bulb globe temperature (WBGTid) ensemble forecasts, both calculated from the corresponding forecasts of temperature and dew point temperature. Two different methodological approaches to calibration are compared. In the first case, we start with joint post-processing of the temperature and dew point forecasts and then create calibrated samples of DI and WBGTid using samples from the obtained bivariate predictive distributions. This approach is compared with direct post-processing of the heat index ensemble forecasts. For this purpose, a novel ensemble model output statistics model based on a generalized extreme value distribution is proposed. The predictive performance of both methods is tested on the operational temperature and dew point ensemble forecasts of the European Centre for Medium-Range Weather Forecasts and the corresponding forecasts of DI and WBGTid. 
For short lead times (up to day 6), both approaches  significantly improve the forecast skill. Among the competing post-processing methods, direct calibration of heat indices exhibits the best predictive performance, very closely followed by the more general approach based on joint calibration of temperature and dew point temperature. 
Additionally, a machine learning approach is tested and shows comparable performance for the case when one is interested only in forecasting heat index warning level categories.

\end{abstract}

\section{Introduction}
\label{sec1}

In this century, extreme temperatures impacted 97 million people causing over 40 billion Euro economic damage \citep{emdat}. In particular, heat stress has detrimental effect on human health and human activities. In order to mitigate the effect of heat stress, dedicated warning systems are developed \citep{morabito2019,mcgregor2015}. These systems require meteorological forecasts from which relevant indicators expressing the impact of heat stress on humans are determined \citep{dnpc19}. A large number of different indices of various complexity exists and even a comparison of 165 different ones demonstrated that there is no universal acceptance of a single index \citep{dfg17}. In this paper, the focus is on two indices  which can be easily derived from standard outputs of weather forecast models and are commonly used: the discomfort index (DI) and the indoor version of the wet-bulb globe temperature (WBGTid). Both are function of temperature and dew point temperature, but have been chosen as they differ in their formulation and complexity which may have an impact on the final findings. 

In our study, forecasts of DI and WBGTid are derived from the ensemble prediction system run at the European Centre for Medium-Range Weather Forecasts (ECMWF). Ensemble predictions provide different scenarios of the future recognising the uncertainties inherit in weather forecasting \citep{lp08}. The derived probabilistic forecasts are an essential tool to support forecast-based decision making \citep{fundel2019}. It has been already established that heat indices forecasts based on ensembles are skillful with respect to climatology up to 10 days \citep{pappenberger2015}. However, the value of these forecast can be increased by correcting the forecasts through so called statistical post-processing taking account of inefficiencies in representing the aforementioned uncertainties and systematic biases. This also adjusts for the fact that the forecasts are produced representing an average over a spatial area (a grid cell) whilst being used and applied locally on individual locations.

Statistical post-processing methods include parametric models such as Bayesian model averaging \citep[BMA;][]{rgbp05} and ensemble model output statistics \citep[EMOS;][]{grwg05}, providing full predictive distributions of the weather variables at hand. We focus on the EMOS approach, which specifies the predictive distribution by a single parametric law with parameters connected to the ensemble members via appropriate link functions. Post-processing methods are often applied to individual weather quantities \citep{wilks18} although the dependence structure between different weather variables is of crucial importance in the calculation of heat stress indicators. 
This is acknowledged and handled if bivariate EMOS models \citep{stg12,bm17} or other approaches to joint calibrations such as the ensemble copula coupling (ECC) method of \citet{stg13} are deployed. More advanced methods \citep[see e.g.][]{bhtp16,barnes19} exist, but are not investigated here.

\begin{figure}[t]
\begin{center}
\epsfig{file=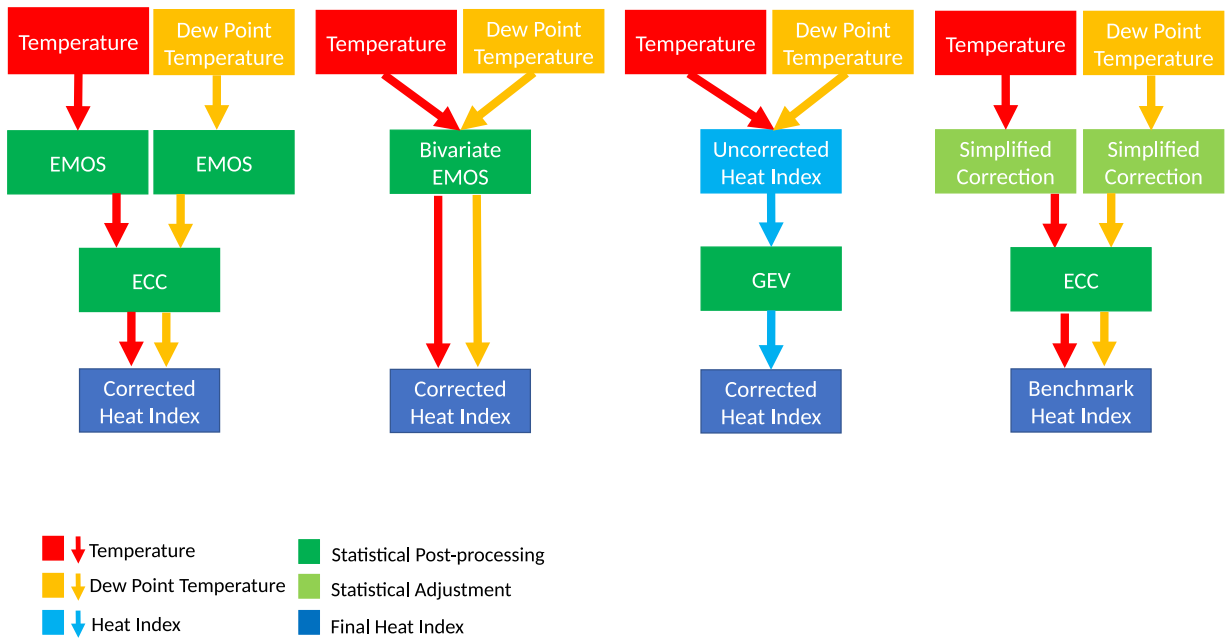, width=.98\textwidth}
\end{center}
\caption{An overview of the different post-processing approaches}
\label{fig:workflow}
\end{figure}

This paper explores the fundamental question whether it is more efficient to base the calibration on the two input forecasts (temperature and dew point temperature) or rather to calibrate directly the end-product (the heat index).
We apply these two different methodological approaches to the calibration of DI and WBGTid ensemble forecasts. On the one hand, we first calibrate ensemble forecast vectors of temperature and dew point temperature, either with a bivariate EMOS model or with ECC combined with independent univariate EMOS models. The calibrated DI and WBGTid forecasts are derived from these calibrated input forecasts. On the other hand, we develop a new EMOS model for post-processing both DI and WBGTid ensemble forecasts, where the predictive PDFs follow a generalized extreme value (GEV) distribution. The benchmark method consists in a simple adjustment of the weather quantities to account for the scale mismatch between model output and point observations.  
A general overview of the applied procedures is given in Figure \ref{fig:workflow}.
The different approaches to post-processing are ranked in terms of performance; their implementation and limitations are also discussed.

Finally, we envisage the situation where one is interested only in the prediction of the warning level category of a heat index. In this case, statistical calibration can be considered as a classification problem resulting in forecast probabilities for the different categories.  Besides obtaining the predictive distributions using the above-mentioned calibration methods, we also test the forecast skill of a general machine learning based approach.

The paper is organized as follows. Section \ref{data} provides the fundamental formulae for the calculation of DI and WBGTid, followed by a description of the temperature and dew point temperature data sets. The applied calibration methods (including the novel GEV EMOS approach) are given in Section \ref{postpro}, and the verification procedure is discussed in Section \ref{verif}. Section \ref{results} presents the results of our study devoting separate sections to bivariate and univariate post-processing, and to  calibration of discrete forecasts. Some figures are placed in the Appendix in order to improve readability. Finally, lessons learned and recommendations can be found in Section \ref{conclusion}.

\section{Data}
\label{data}

\subsection{Heat index definitions}
\label{heatindices}

We start with the formal definition of the two heat indices of interest, which are both functions of two weather quantities:  temperature and dew point (DP) temperature. Discomfort index \ $I_d$ \ is  calculated from temperature \ $T$ \ ($^\circ$C)  and relative humidity (RH) \ $H_r$  \ ($\%$) as
\begin{equation}
 \label{eq:DI}
I_d:= T - 0.0055(100-H_r)(T-14.5). 
\end{equation}
RH can easily be expressed using temperature and DP temperature \ $T_{dp}$ \ ($^\circ$C) using Magnus formula with constants suggested by \citet{sonntag90}
\begin{equation}
  \label{eq:RH}
  H_r := 100 \exp \bigg(\frac {17.62\,T_{dp} }{243.12 + T_{dp}} - \frac {17.62\,T}{243.12 + T} \bigg).
\end{equation}

Indoor wet-bulb globe temperature \ $T_{WBGid}$ \ is calculated as
\begin{equation}
 \label{eq:WBGTid}
	T_{WBGid}:= 0.67\,T_{pwb} + 0.33T,
\end{equation}
where \ $T_{pwb}$ \ denotes the psychrometric wet-bulb (PWB) temperature ($^\circ$C). One can obtain this quantity using Bernard's formula \citep{bp99} providing it as a solution of equation
\begin{equation*}
1556\, P_d -1.484\,P_d \,T_{pwb} - 1556\, P_w + 1.484\,P_w \,T_{pwb} + 1010\,(T - T_{pwb}) = 0,
\end{equation*}
where
\begin{equation*}
P_d = 6.106 \exp \bigg(\frac{17.27\, T_{dp}}{237.3+T_{dp}}\bigg) \qquad \text{and} \qquad
P_w = 6.106 \exp \bigg(\frac{17.27\, T_{pwb}}{237.3+T_{pwb}}\bigg) 
\end{equation*}
are the saturation vapour pressures (hPa) at DP temperature \ $T_{dp}$ \ and PWB temperature \ $T_{pwb}$, \ respectively \citep{lk12}. It results that WBGTid is also a function of temperature and dew point and can be computed using e.g. the {\tt HeatStress} package of {\tt R} \citep{hs19}.

\subsection{Observations}

\begin{figure}[t]
\begin{center}
\epsfig{file=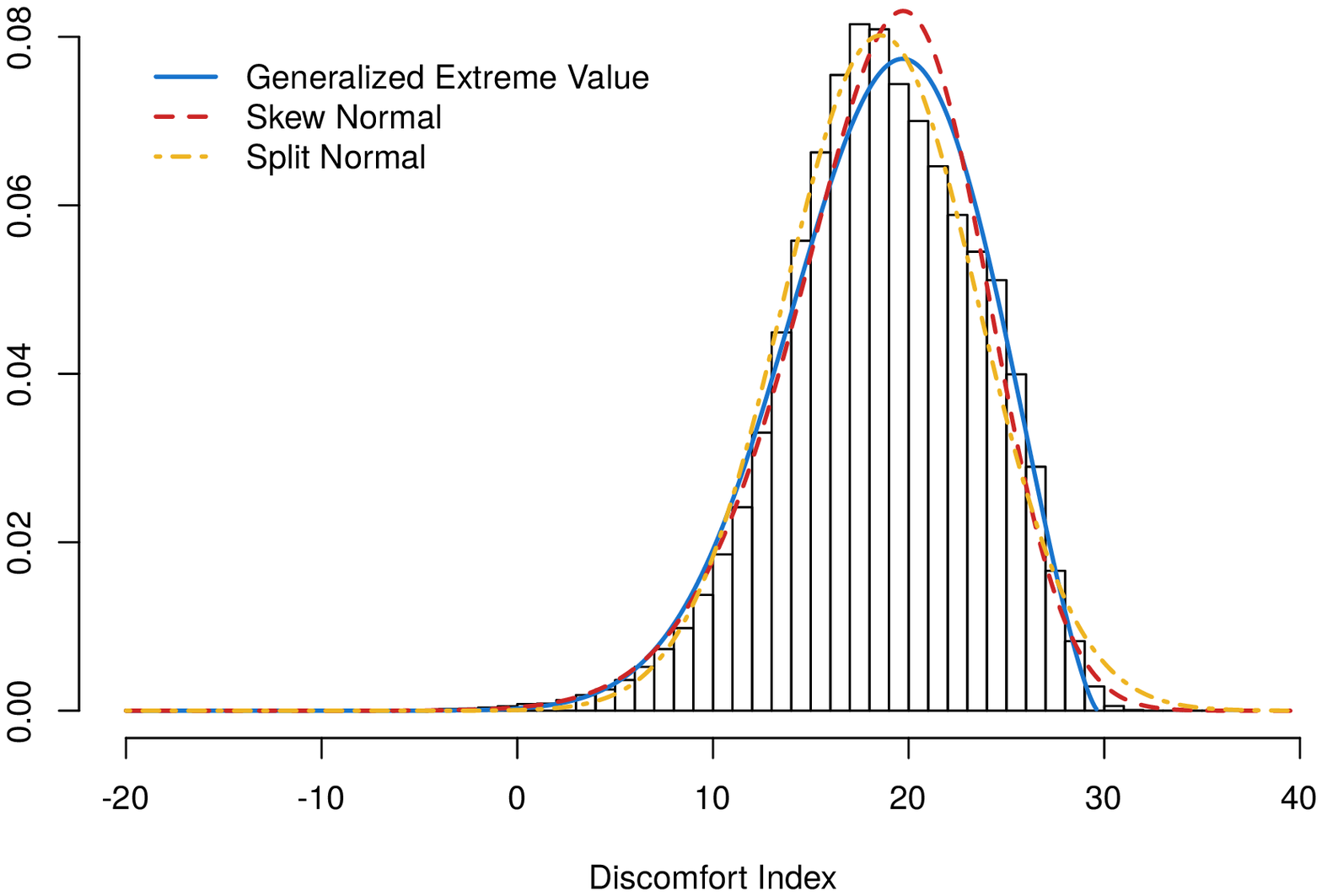, width=.47\textwidth} \qquad
\epsfig{file=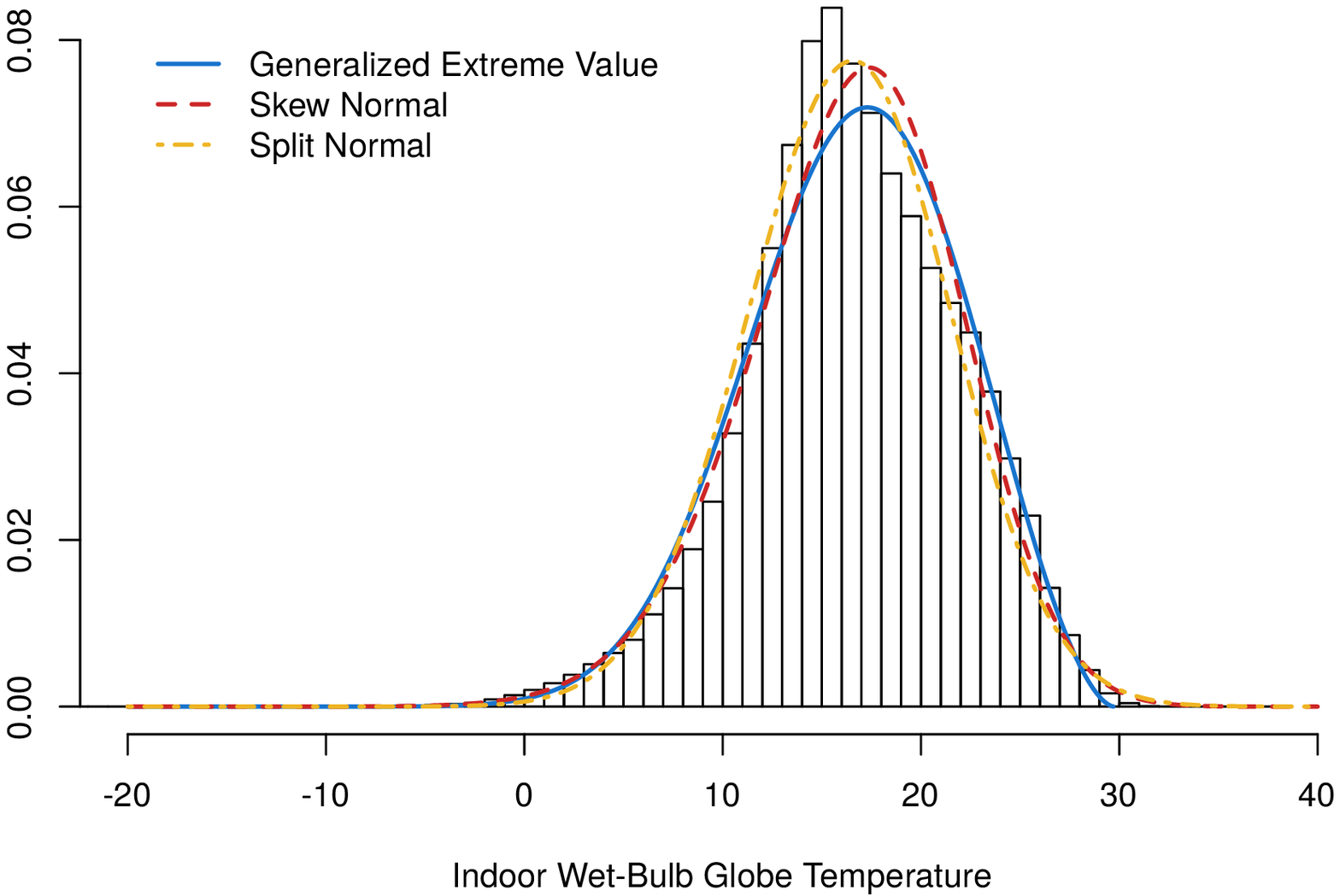, width=.47\textwidth} 
\end{center}
\caption{Climatological histograms of DI ({\em left}) and WBGTid ({\em right}) and the PDFs of generalized extreme value, skew normal and split normal distributions with matching means, variances and skewnesses.}
\label{fig:climHists}
\end{figure}

Temperature and dew point observations are available at 1459 surface synoptic observation (SYNOP) stations in the mid-latitudes of Europe (latitudes: 35$^\circ$N -- 65$^\circ$N; longitudes: 12.5$^\circ$W -- 42.5$^\circ$E).
The corresponding DI is derived by applying \eqref{eq:DI} and \eqref{eq:RH}, whereas WBGTid is obtained from \eqref{eq:WBGTid}, where PWB is determined with the help of the {\tt HeatStress} package. The selected data set covers a period between 1 May and 30 September 2017.

Both DI and WBGTid observations are negatively skewed (skewnesses are \ $-0.415$ \ and \ $-0.286$, \ respectively). This should be taken into account in the choice of the parametric distribution family for EMOS post-processing. The left-skewed character of both indices is visible in Figure \ref{fig:climHists} showing the corresponding climatological histograms together with the PDFs of generalized extreme value (GEV), skew normal and split normal distributions (with matching means, variances and skewnesses). These three parametric distributions are potential candidates for  EMOS modeling (see Section   \ref{GEVEMOSmodel}).

\subsection{Ensemble forecasts}

The operational ensemble run at ECMWF comprises 50 perturbed members. We focus on forecasts initialized at 12 UTC and we consider 1 - 15 day lead times.  For each observation, we associate the forecast at the nearest grid point. In order to account for the difference between model and station elevations, orographic correction is applied  to both raw temperature and dew point forecasts. Practically, we add \ $ 0.0065  \Delta_{e}$ \ ($^\circ$C) to the forecasts, with \ $\Delta_{e}$ \ the altitude difference between station and model representation.  Temperature and dew point forecasts are used to derive the corresponding raw ensemble forecasts of DI and WBGTid. 

Raw forecasts provide information on a grid (here the model-grid resolution is $\sim18$ km), whereas observations are point measurements. This scale mismatch leads to representativeness error in the forecast. The raw forecast uncertainty, valid at the model-resolution scale, can be increased in order to capture the temperature variability at smaller spatial scale. The method followed here is a simple down-scaling step, inspired by the pioneer work of \cite{saetra2004}: it consists in adding to each member a draw from a centered Gaussian distribution with standard deviation
\begin{equation}
\sigma_{pert}:= 0.75 + 0.18 \, \big|\Delta_{e} \big|^{1/4}.
\label{adjustment}
\end{equation}
This formula is derived from the analysis of 2 m temperature measurements  of a high-density observation network \citep{zbb2020}. The spread adjustment is applied independently to raw temperature and dew point ensemble forecasts, followed by a reordering of the adjusted forecasts using the rank of the raw forecasts before the calculation of DI and WBGTid forecasts (see Figure \ref{fig:workflow}).  These corrected forecasts are referred to as the adjusted ensemble and are used as a reference for the computation of skill scores (see Section \ref{skillscores}).

\section{Calibration methods}
\label{postpro}

Our main approach to calibration is the computationally efficient EMOS method that has demonstrated excellent predictive performance for a wide range of weather quantities. In what follows, we denote \ $f_1,f_2, \ldots ,f_K$ \ the ensemble forecast of a given weather quantity for a given location, time and lead time, and  $\overline f$ \ the ensemble mean. \ $S^2$ \ and \ $\md$ \ denote the ensemble variance and the ensemble mean absolute difference, respectively, defined as 
\begin{equation*}
  S^2 := \frac 1{K-1} \sum_{k=1}^K \big(f_k - \overline f\big)^2 \qquad \text{and} \qquad \md := \frac 1{K^2}\sum_{k,\ell =1}^K \big|f_k - f_{\ell}\big|.
\end{equation*}
Multivariate ensemble forecasts are distinguished by bold notation \ $\boldsymbol f_1,\boldsymbol f_2, \ldots ,\boldsymbol f_K$, \ and, in this case, the ensemble variance is replaced by the ensemble covariance matrix 
\begin{equation*}
\boldsymbol S^2 := \frac 1{K-1} \sum_{k=1}^K \big(\boldsymbol f_k - \overline{\boldsymbol f}\big)\big(\boldsymbol f_k - \overline{\boldsymbol f}\big)^{\top}.
\end{equation*}

The 50 members of the operational ECMWF ensemble system are considered as statistically indistinguishable and, in this sense, exchangeable. In the following, if we have \ $M$ \ ensemble members divided into \ $K$ \ exchangeable groups, where the $k$th group contains \ $M_k\geq 1$ \ ensemble members \ ($\sum_{k=1}^KM_k =M$), \  notation \ $\overline f_k$ \ ($\overline {\boldsymbol f}_k$) \ will be used for the corresponding group mean (vector).

\subsection{Normal EMOS model}
\label{normalEMOSmodel}
  
Calibration of temperature and dew point ensemble forecasts is based on the EMOS model suggested by \citet{grwg05}. The predictive distribution is a Gaussian distribution \ ${\mathcal N}(\mu,\sigma ^2)$ \ with mean \ $\mu$ \ and variance \ $\sigma^2$ \ linked to the ensemble members via equations
\begin{equation}
  \label{eq:normEmos1D}
 	\mu = a +b_1 f_1 + \cdots + b_K f_K \qquad \text{and} \qquad  \sigma^2 := c^2 + d^2 S^2,
\end{equation}
where \ $a, b_1, \ldots , b_K, c, d \in {\mathbb R}$. \ When an ensemble contains groups of statistically indistinguishable ensemble members, members within a given group should share the same parameters \citep{gneiting14, wilks18} and the equation for the mean in \eqref{eq:normEmos1D} is replaced by
\begin{equation}
  \label{eq:normEmos1Dex}
 	\mu = a +b_1 \overline f_1 + \cdots + b_K \overline f_K. 
\end{equation}

\subsection{Bivariate normal EMOS model}
  \label{BivariateEMOSmodel}
The separate univariate EMOS calibration of temperature and dew point ensemble forecasts does not take into account the obvious dependence between these two weather quantities. In order to overcome  this limitation, one can consider the joint post-processing of these two variables using an EMOS model with a bivariate normal predictive distribution \ ${\mathcal N}_2\big(\boldsymbol\mu, \boldsymbol\Sigma)$, \ where the mean vector \ $\boldsymbol \mu$ \ and covariance matrix \ $\boldsymbol\Sigma$ \ are expressed as
\begin{equation}
   \label{eq:normEMOS2D}
 \boldsymbol \mu =  {\mathcal A}+{\mathcal B}_1\boldsymbol f_1+ \cdots
  +{\mathcal B}_K \boldsymbol f_K \qquad \text{and} \qquad \boldsymbol\Sigma = {\mathcal C}{\mathcal C}^{\top}+{\mathcal D}{\boldsymbol S}^2 {\mathcal D}^{\top},
\end{equation}
respectively. Here  \ ${\mathcal A}\in {\mathbb R}^2$, \ whereas \ ${\mathcal B}_1, \ldots , {\mathcal B}_K$ \ and \ ${\mathcal C},\ {\mathcal D}$ \ are two-by-two real parameter matrices, \ where \ $\mathcal C$ \ is lower triangular. This way, \ ${\mathcal C}{\mathcal C}^{\top}$ \ is  symmetric and non-negative definite. 
In the case of existence of groups of exchangeable ensemble members, the bivariate generalization of \eqref{eq:normEmos1Dex} is
\begin{equation}
   \label{eq:normEMOS2Dex}
 \boldsymbol \mu =  {\mathcal A}+{\mathcal B}_1\overline{\boldsymbol f}_1+ \cdots
  +{\mathcal B}_K \overline{\boldsymbol f}_K. 
\end{equation}
Note that a bivariate EMOS model with a different parametrization of the covariance matrix has already been successfully applied for the calibration of wind vector ensemble forecasts \citep{stg12}.

\subsection{Generalized extreme value EMOS model}
  \label{GEVEMOSmodel}

In contrast to temperature and dew point (where the symmetric Gaussian distribution provides a reasonable fit), both DI and WBGTid observation distributions result in left-skewed histograms, calling for predictive PDFs with negative skewness. Potential candidates are the following: a skew normal, a split normal or a GEV distribution (see  Figure \ref{fig:climHists}). For these three competing laws, the GEV seems to best capture the tail behaviour of the observations (at least for the DI and WBGTid observations at hand).

A GEV distribution \ $\mathcal{GEV}\big(\mu,\sigma,\xi\big)$ \ with location \ $\mu $, \ scale \ $\sigma>0$ \ and shape \ $\xi$ \ can be characterized by its cumulative distribution function (CDF)
\begin{equation}
    \label{eq:gevCDF}
    H(x\vert\, \mu,\sigma,\xi ):=\begin{cases}
\exp\Big(-\big[1+\xi(\frac{x-\mu}{\sigma})\big]^{-1/\xi}\Big), & \
\xi\ne 0; \\
\exp\Big(-\exp\big(-\frac{x-\mu}{\sigma}\big)\Big), & \
\xi= 0,
\end{cases} 
\end{equation}  
if \ $1+\xi(x-\mu)/\sigma> 0$ \ and zero otherwise, which for \ $\xi <1$ \ has a finite mean. For calibrating ensemble forecasts of DI and WBGTid, we suggest modeling location and scale parameters by 
\begin{equation}
	\label{eq:gevEMOS}
	\mu = \alpha +\beta_1 f_1 + \cdots + \beta_K f_K \qquad \text{and} \qquad  \sigma := \gamma^2 + \delta^2\, \md
\end{equation}
with \ $\alpha,\beta_1, \ldots, \beta_K, \gamma, \delta \in {\mathbb R}$, \ whereas the shape parameter \ $\xi$ \ is considered to be independent of the ensemble members. Similar to \eqref{eq:normEmos1Dex} and \eqref{eq:normEMOS2Dex}, the exchangeable version of \eqref{eq:gevEMOS} considers the location \ $\mu$ \ as an affine function of the group means:
\begin{equation}
	\label{eq:gevEMOSex}
	\mu = \alpha +\beta_1 \overline f_1 + \cdots + \beta_K \overline f_K.
\end{equation}
The expression of the mean (or location) as an affine function of the ensemble is a general feature in EMOS post-processing regardless of the weather quantity at hand \citep[see e.g.][]{grwg05,tg10,bn16}, whereas the dependence of the scale on the ensemble mean difference is similar to the censored GEV model of \citet{sch14} for calibrating precipitation ensemble forecasts. Exploratory tests with the data set at hand show that the proposed model significantly outperforms the GEV EMOS model with scale depending on the ensemble mean, that is when \ $\sigma = \gamma^2 + \delta^2 \overline f$ \ \citep[see e.g.][]{lt13}, and results in slightly (and not always significantly) better predictive performance than its natural modifications involving the ensemble variance \ $S^2$, \ such as
\begin{equation*}
\sigma = \sqrt{\gamma^2 + \delta^2 S^2} \qquad \text{or} \qquad \sigma = \gamma^2 + \delta^2 \sqrt{S^2}.
\end{equation*}

\subsection{Parameter estimation}
  \label{paramEstim}

Parameters of the normal and GEV EMOS models described in Sections \ref{normalEMOSmodel} -- \ref{GEVEMOSmodel} are estimated according to the optimum score estimation principle of \citet{gr07}. The estimates of the unknown parameters are obtained by optimizing the mean value of a proper scoring rule over an appropriate training data-set. The most popular proper scoring rules are the logarithmic (or ignorance) score \citep[LogS;][]{good52}, that is the negative logarithm of the predictive PDF at the verifying observation and the continuous ranked probability score \citep[CRPS;][]{gr07,wilks11}. The CRPS corresponding to a (predictive) CDF \ $F(y)$ \ and a real value (observation) \ $x$ \ is defined as
\begin{equation}
  \label{eq:CRPS}
\crps\big(F,x\big):=\int_{-\infty}^{\infty}\big (F(y)-{\mathbb 
  I}_{\{y \geq x\}}\big )^2{\mathrm d}y={\mathsf E}|X-x|-\frac 12
{\mathsf E}|X-X'|, 
\end{equation}
where \ ${\mathbb I}_H$ \ denotes the indicator of a set \ $H$, \ whereas \ $X$ \ and \ $X'$ \ are independent random variables with CDF \ $F$ \ and finite first moment. The last representation in \eqref{eq:CRPS} implies that the CRPS can be expressed in the same unit as the observation. Both CRPS and LogS are negatively oriented scoring rules (the smaller the better), and the optimization with respect to LogS  gives back the maximum likelihood (ML) estimation of the parameters. Furthermore, the CRPS for both normal and GEV distributions can be expressed in closed form (for the exact formulae see \citet{tg10} and \citet{ft12}, respectively), thereby allowing for efficient parameter estimation procedures often being more robust than the ML approach.

The logarithmic score is obviously well defined also in the case of multivariate predictive distributions. Further, the direct multivariate extension of the CRPS is the energy score (ES) introduced by \citet{gr07}.  Given a CDF \ $F$ \ on \ ${\mathbb R}^d$ \ and a $d$-dimensional vector \ $\boldsymbol x$, \ the energy score is defined as
\begin{equation}
 \label{eq:ES}
\es(F,\boldsymbol x):={\mathsf E}\Vert \boldsymbol X-\boldsymbol
x\Vert-\frac 12 {\mathsf E}\Vert \boldsymbol X-\boldsymbol X'\Vert,
 \end{equation}
where \ $\Vert \cdot \Vert$ \ denotes the Euclidean distance and, similar to the univariate case, \ $ \boldsymbol X$ \ and \  $\boldsymbol X'$ \ are independent random vectors having distribution \ $F$. \ In most cases (including the case of the Gaussian model), ES cannot be given in a closed form, so it should be replaced by a Monte Carlo approximation based on a large sample drawn from \ $F$ \ \citep{gsghj08}. In our study, we apply ML estimation for bivariate models, whereas parameters of all univariate EMOS models are obtained by optimizing the CRPS. In the latter case, tests using  ML estimation have not demonstrated  significantly better results.

The appropriate choice of training data is also an important issue in statistical post-processing. In EMOS modelling, the standard approach is the use of rolling training periods where the parameter estimates are obtained using ensemble forecasts and corresponding validating observations for the preceding \ $n$ \ calendar days. Once the appropriate training period length is chosen, there are two general approaches to spatial selection of training data \citep{tg10}. In the local approach, model parameters are estimated using only training data of the given station, resulting in distinct parameter estimates for the different stations. The global (regional) approach uses training data of the whole ensemble domain and all observation stations share the same set of EMOS parameters. Global modelling requires a far shorter training period but is usually unsuitable for large and heterogeneous data-sets like the one at hand. Alternatively, one can combine the advantages of local and global parameter estimation using a semi-local approach \citep{lb17}: the training data for a given station is enhanced with data from stations with similar characteristics. Following \citet{blszbb19}, clustering based semi-local modelling
where regional parameter estimation is also considered. The clusters are derived using $k$-means clustering of feature vectors depending both on the station climatology and the forecast errors of the raw ensemble mean over the training period.

\subsection{Ensemble copula coupling}
  \label{ECC}

The ensemble copula coupling (ECC) approach of \citet{stg13} is a general multivariate post-processing method, where after univariate calibration, the rank order information in the raw ensemble is used to restore correlations between the calibrated forecasts. In our case, we first perform univariate EMOS calibration of temperature and dew point forecasts using the normal EMOS model of Section \ref{normalEMOSmodel}. Then we draw \ $N$ \ random samples of the same size as the raw ensemble from both individual EMOS predictive distributions \citep[ECC-R; see e.g.][]{schefzik16} and reorder them to have the same rankings as the corresponding raw temperature and dew point forecasts. The obtained bivariate sample is considered as a sample from a calibrated bivariate predictive distribution of temperature and dew point.

In practice, when applying the ECC approach, 20 samples of the size of the raw ensemble (50) are drawn from the EMOS predictive distributions of temperature and dew point. Each sample is then reordered according to the rank structure of the corresponding raw ensemble forecasts resulting in 1000 draws from the ECC predictive distribution, which can be used for forecast evaluation. In a similar way, verification scores of the bivariate EMOS models are also estimated using 1000 samples from the bivariate predictive distributions. 

\section{Verification}
\label{verif}
  
\subsection{Proper scores}
\label{scores}
For a given lead time,  competing forecasts in terms of probability distribution are compared with the help of the mean CRPS in the univariate case and mean ES value in the bivariate case over all forecast cases in the verification data. The skill of univariate forecasts for dichotomous events (observation  \ $x$ \ exceeding a given threshold \ $y$) \ is assessed with the mean Brier score (BS). \ For a predictive CDF \ $F(y)$, \ the Brier score is defined as
\begin{equation*}
 \bs \big(F,x;y\big):= \big (F(y)-{\mathbb I}_{\{y \geq x\}}\big )^2.
\end{equation*}
The integral of BS over all possible thresholds results in the CRPS. The goodness of fit of probabilistic forecasts to extreme values of the univariate weather quantity  is evaluated using the threshold-weighted continuous ranked probability score (twCRPS) 
\begin{equation*}
\twcrps\big(F,x\big):=\int_{-\infty}^{\infty}\big (F(y)-{\mathbb 
  I}_{\{y \geq x\}}\big )^2\omega(y){\mathrm d}y,
\end{equation*}
where \ $\omega(y)\geq 0$ \ is a weight function \citep{gr11}. The case where \ $\omega(y)\equiv 1$ \ corresponds to the CRPS defined in  \eqref{eq:CRPS}. In the following, we set \ $\omega(y)={\mathbb I}_{\{y \geq r\}}$ in order to focus exclusively on values above a given threshold \ $r$. \ 

\subsection{Skill scores}
\label{skillscores}

The improvement in terms of  a given score \ ${\mathcal S}_F$ \ for a forecast \ $F$ \ with respect to a reference forecast \ $F_{ref}$ \  can be quantified using the corresponding skill score
\begin{equation}
\label{eq:skillScore}
  {\mathcal S}^{skill}_F :=\frac{\overline{\mathcal S}_{F_{ref}} - \overline{\mathcal S}_F}{\overline{\mathcal S}_{F_{ref}} - \overline{\mathcal S}_{F_{perf}} }, 
\end{equation}
where \ $\overline{\mathcal S}_{F}$ \ and \ $\overline{\mathcal S}_{F_{ref}}$  \ denote the mean score values corresponding to forecasts \ $F$ \ and  \ $F_{ref}$, \ respectively, and \ $\overline{\mathcal S}_{F_{perf}}$  \ corresponds to the mean score value of a perfect deterministic forecast (with here \ $\overline{\mathcal S}_{F_{perf}}  =0 $). \  Based on \eqref{eq:skillScore},  we compute the continuous ranked probability skill score (CRPSS), the energy skill score (ESS), the Brier skill score (BSS) and the threshold-weighted continuous ranked probability skill score (twCRPSS). Skill scores are positively oriented (the larger the better), and in Section \ref{results} the adjusted ensemble is used as a reference.

\subsection{Calibration assessment}
\label{calibration}

A simple graphical tool for assessing calibration of univariate ensemble forecasts is the verification rank histogram, which is the histogram of ranks of verifying observations with respect to the corresponding ensemble forecasts \citep[see e.g.][Section 8.7.2]{wilks11}. In the case of a perfectly calibrated $K$-member ensemble, the ranks follow a uniform distribution on \ $\{1,2,\ldots ,K+1\},$ \ and the deviation from uniformity can be quantified by the reliability index \ $\Delta$ \ defined by
\begin{equation}
   \label{eq:relind}
 \Delta:=\sum_{r=1}^{K+1}\Big| \rho_r-\frac 1{K+1}\Big|,
\end{equation}
where \ $\rho_r$ \ is the relative frequency of rank \ $r$ \ \citep{delle}.  One can also generalize the verification rank histogram to multivariate ensemble forecasts. \citet{tsh16} suggested several options in order to define ranks in a multivariate context. We apply the average ranking given by the average of the univariate ranks and the multivariate ordering proposed by \citet{gsghj08}. In case of a probabilistic forecast, one can plot the verification rank histogram and  calculate corresponding reliability index on the basis of a large number of ensembles sampled from the predictive PDF. 

Calibration assessment of univariate predictive distributions is examined with the help of probability integral transform (PIT) histograms (the continuous counterparts of the verification rank histograms). By definition, PIT is the value of the predictive CDF at the validating observation, with a possible randomization at points of discontinuity \citep{gr13}. In case of perfect calibration, PIT follows a uniform law on the \ $[0,1]$ \ interval.

\subsection{Forecast value}
\label{value}

The forecast value compares mean expenses of users having  access to different levels of information; it has the form of a skill score  (see Section \ref{skillscores}). The framework is the one of a simple binary decision setting: facing a potential threatening event (i.e. a heat index exceeding a warning level), a user can take (or not) preventive action. This protective action has a cost \ $C$, \ whereas the user can endure a loss \ $L$ \ if the event actually occurs but no protective action was taken anticipatively. The ratio \ $C/L$ \ is called cost-loss ratio and is considered as a characteristic of the user in the decision-making process.
The user takes action based on the forecast: if the forecast probability is greater than the user's cost-loss ratio then preventive action with cost \ $C$ \ is taken, otherwise the risk to face a loss \ $L$ \  in case of event is taken. This rational behaviour is sometimes referred to as the Bayes action (see \citet{rodwell2019} for a discussion on the concept and applications). When only climatological information is available, the user's cost-loss ratio is simply compared to the event base rate. In this framework, the forecast value is defined as the ratio of two mean expense differences involving the following elements: the mean expense of the user taking the Bayes action with respect to the forecast \ ($\overline{E}_{F})$, \  the mean expense of a user making decision based on the observation climatology \ ($\overline{E}_{F_{clim}}$), \ and the mean expense of an omniscient user who is taking action only when the event actually occurs \ ($\overline{E}_{F_{perf}}$). \ Formally, the forecast value is defined as
\begin{equation*}
V = \frac{\overline{E}_{F} -\overline{E}_{F_{clim}}} {\overline{E}_{F_{perf}} -\overline{E}_{F_{clim}}},
\end{equation*}
which can be expressed in a simple form using the elements of a contingency table \citep[see e.g.][]{wilks01}. In our implementation, the mean expenses of the different type of users are computed independently at each station. This means that we consider a climatology (estimated as the sample base rate) varying from one station to another.

\section{Results}
\label{results}
  
We consider two different methodological approaches to the calibration of DI and WBGTid. In the first case, we start with joint post-processing of temperature and dew point ensemble forecasts, either using directly a bivariate normal EMOS model or applying first univariate normal EMOS models to both components and then ECC to restore the dependence structure. Samples from the bivariate predictive distributions are then used to obtain samples from the predictive distributions of DI and WBGTid. The second approach is the direct post-processing of DI and WBGTid ensemble forecasts with the help of the novel GEV model introduced in Section \ref{GEVEMOSmodel} (see Figure \ref{fig:workflow}).

\subsection{Bivariate calibration of temperature and dew point forecasts}
\label{bivariatecalib}

We start with some considerations about the implementation of the EMOS approaches. The 50 ensemble members are regarded as exchangeable; their separate univariate and joint bivariate calibration is performed using normal EMOS models with distribution means linked to the ensemble means via \eqref{eq:normEmos1Dex} and \eqref{eq:normEMOS2Dex}, respectively, with \ $K=1$. \ Thus, the EMOS models require the estimation of  $4$ free parameters in the univariate case and  $13$ in the bivariate case. A large number of free parameters calls either for long training periods in local estimation or for a semi-local approach.  Both local and clustering-based semi-local parameter estimations with 5, 10, 15, 25, 50 and 75 clusters are tested.  Clustering of stations is based on 40 dimensional features equally representing the properties of both temperature and dew point forecasts and observations. In accordance with the results of \citet{lb17}, the forecast skill does not really depend on the number of features applied, provided that one works with a sufficiently large number of features. Moreover, the longer the lead time of the post-processed forecasts, the more training data are needed to outperform the raw ensemble forecasts \citep[see discussion in][]{frg19}.  As a trade-off, the length of the  training period is fixed to 60 days. This choice leaves 79 calendar days (period 14 July -- 30 September 2017) as a verification period when taking into account the maximal lead time of 15 days.

\begin{figure}[t]
\begin{center}
\epsfig{file=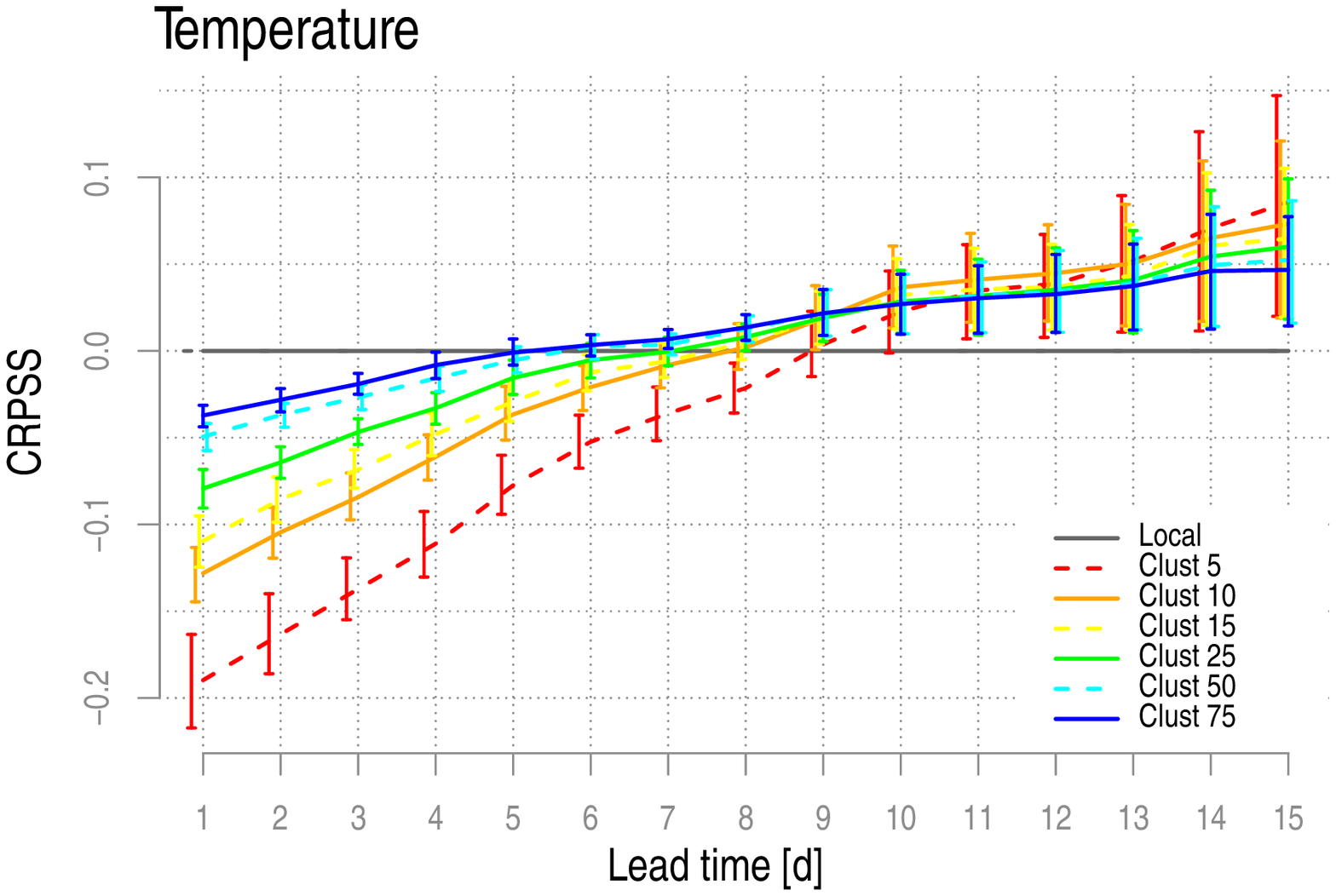, width=.47\textwidth} \qquad
\epsfig{file=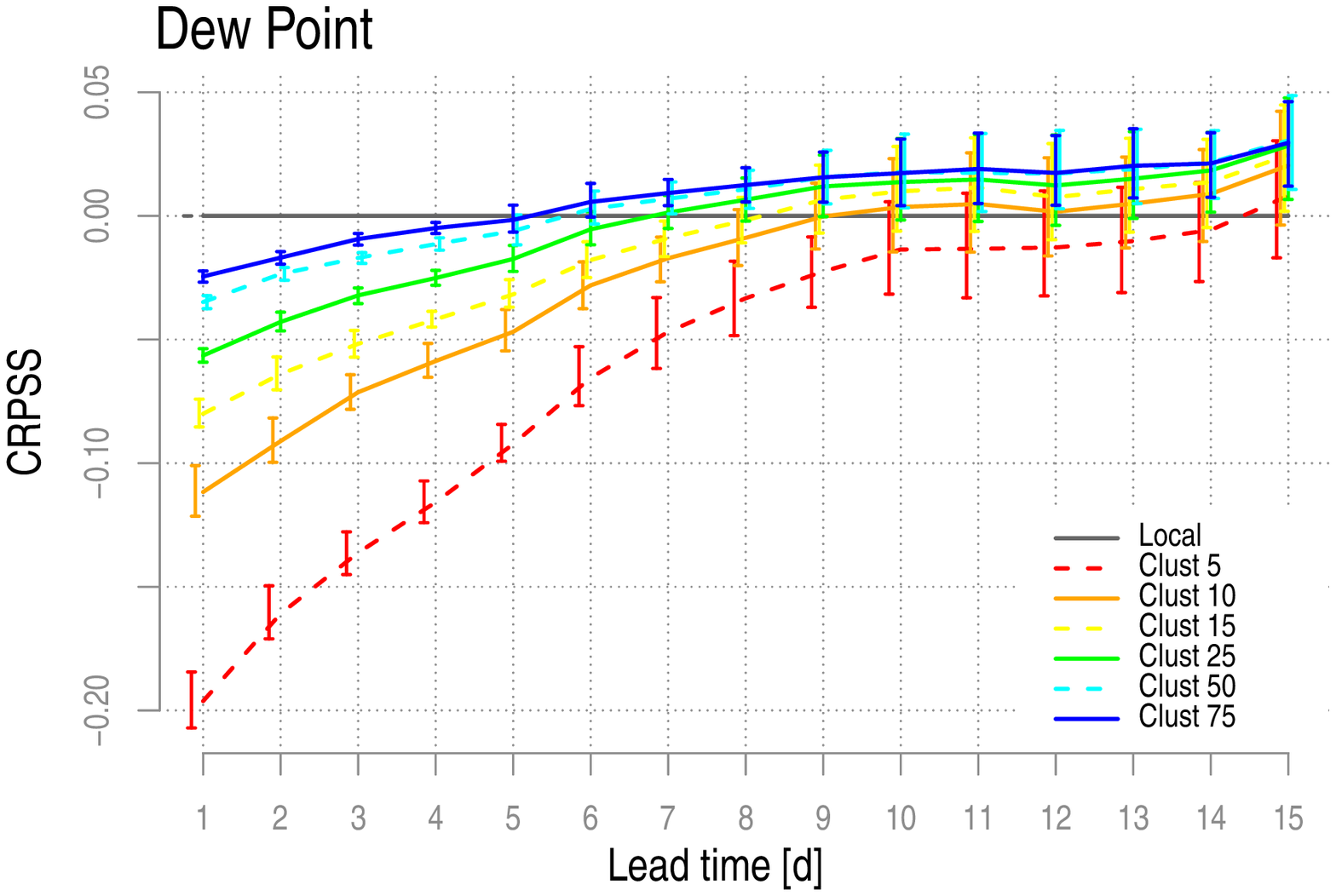, width=.47\textwidth} 
\end{center}
\caption{CRPSS of semi-local EMOS models with respect to the local EMOS for temperature ({\em left}) and dew point ({\em right}) together with $95\,\%$ confidence intervals.}
\label{fig:crpssTempDP}
\end{figure}

First,  calibration of temperature and dew point forecast is performed separately. Figure \ref{fig:crpssTempDP} shows the CRPSS of different semi-local EMOS models with respect to local EMOS. Local post-processing is optimal up to day 5, then semi-local approaches exhibit significantly better predictive performance. 
Hence, we propose to apply different types of EMOS configurations for different forecast lead times: 
local EMOS at the beginning of the forecast range and semi-local configuration(s) for longer lead times.
Practically, local EMOS is considered for days 1 -- 5 both for temperature and dew point, while for days 6 -- 10 we use semi-local EMOS with 75 clusters. For dew point, this latter model is also kept for days 11 -- 15, whereas for temperature we reduce the number of clusters to 5. The different mixtures of EMOS configurations are reported in  Table \ref{tab:tab1}. In the following, the method referred to as {\em ECC\/} relies on the reported mixed approach for the univariate post-processing step.

\begin{table}[t]
\begin{center}{\footnotesize
\begin{tabular}{l|c|c|c|c|c|c|c|c|c|c|c|c|c|c|c|}
\multicolumn{1}{l|}{Post-processing}&\multicolumn{15}{c}{Lead time (day)}\\ \cline{2-16}
method&1&2&3&4&5&6&7&8&9&10&11&12&13&14&15\\ \hline
Normal EMOS, temperature&\multicolumn{5}{c|}{local}&\multicolumn{5}{c|}{75 clusters}&\multicolumn{5}{c|}{5 clusters} \\ \hline
Normal EMOS, dew point&\multicolumn{5}{c|}{local}&\multicolumn{10}{c|}{75 clusters} \\ \hline
2D EMOS &\multicolumn{4}{c|}{local}&\multicolumn{3}{c|}{50 clusters}&\multicolumn{8}{c|}{5 clusters} \\ \hline
2D EMOS optimal&\multicolumn{5}{c|}{local}&\multicolumn{5}{c|}{75 clusters}&\multicolumn{5}{c|}{5 clusters} \\ \hline
GEV EMOS, DI and WBGTid&\multicolumn{4}{c|}{local}&\multicolumn{4}{c|}{75 clusters}&\multicolumn{7}{c|}{5 clusters} \\ \hline
\end{tabular}
}
\end{center}
\caption{Training configurations of EMOS models for different forecast lead times. \label{tab:tab1}}
\end{table}

In the same spirit as Figure \ref{fig:crpssTempDP}, the optimal training configuration is investigated by inspecting the multivariate performance in terms of ESS for the semi-local bivariate EMOS models with respect to the local one. As reported in Table \ref{tab:tab1}, optimal performance in terms of ES can be obtained by applying local modelling for days 1 -- 5, semi-local with 75 clusters for days 10 -- 15, and semi-local with 5 clusters for longer lead times (referred to as {\em 2D EMOS optimal\/}). However, this mixture of EMOS configurations is sub-optimal when applied to derive predictive distributions of DI and WBGTid 
(see Section \ref{calibDIWBGT}). Better forecast skill can be obtained using local EMOS for days 1 -- 4, semi-local with 50 clusters for days 5 -- 7, and semi-local with 5 clusters until the end of the forecast range (referred to as {\em 2D EMOS\/}).

\begin{figure}[t]
\begin{center}
\epsfig{file=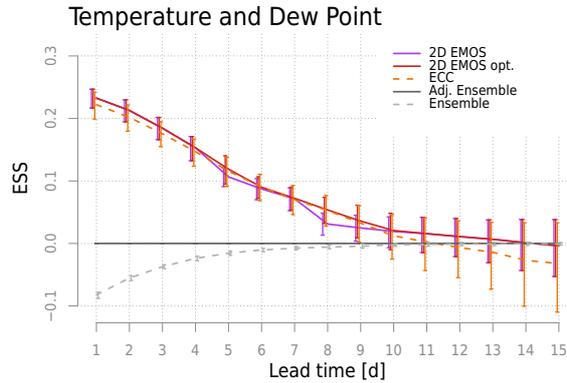, width=.47\textwidth} 
\end{center}
\caption{ESS with respect to the adjusted ensemble of bivariate forecasts together with $95\,\%$ confidence intervals.}
\label{fig:es2D}
\end{figure}

In Figure \ref{fig:es2D}, the different multivariate post-processing approaches are compared in terms of a multivariate skill score, ESS.  Bivariate EMOS and ECC post-processing approaches significantly outperform both the raw ensemble and the adjusted ensemble up to day 9, while the advantage of the adjusted ensemble with respect to the raw ensemble disappears only after day 10. All  performance differences shrink with forecast lead time. Bivariate EMOS generally outperforms ECC but both methods have similar performance between day 5 and day 10. When significant difference in mean ES between these two approaches is tested, it appears that the proportion of stations with significant difference (positive or negative) is less than 7\% at day 1 but increases to more than 30\% at day 10 (not shown).

\begin{figure}[t]
\begin{center}
\epsfig{file=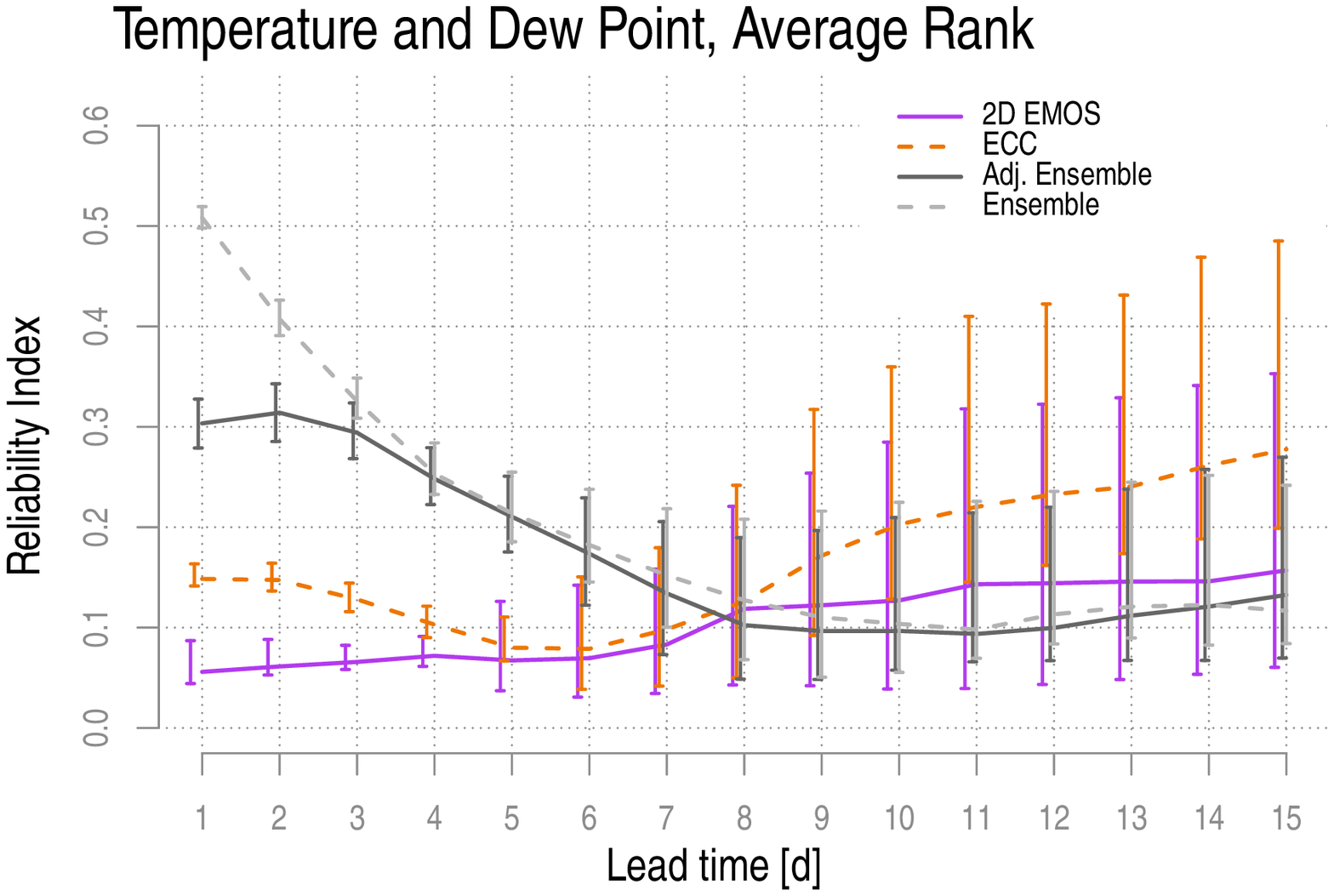, width=.47\textwidth} \qquad
\epsfig{file=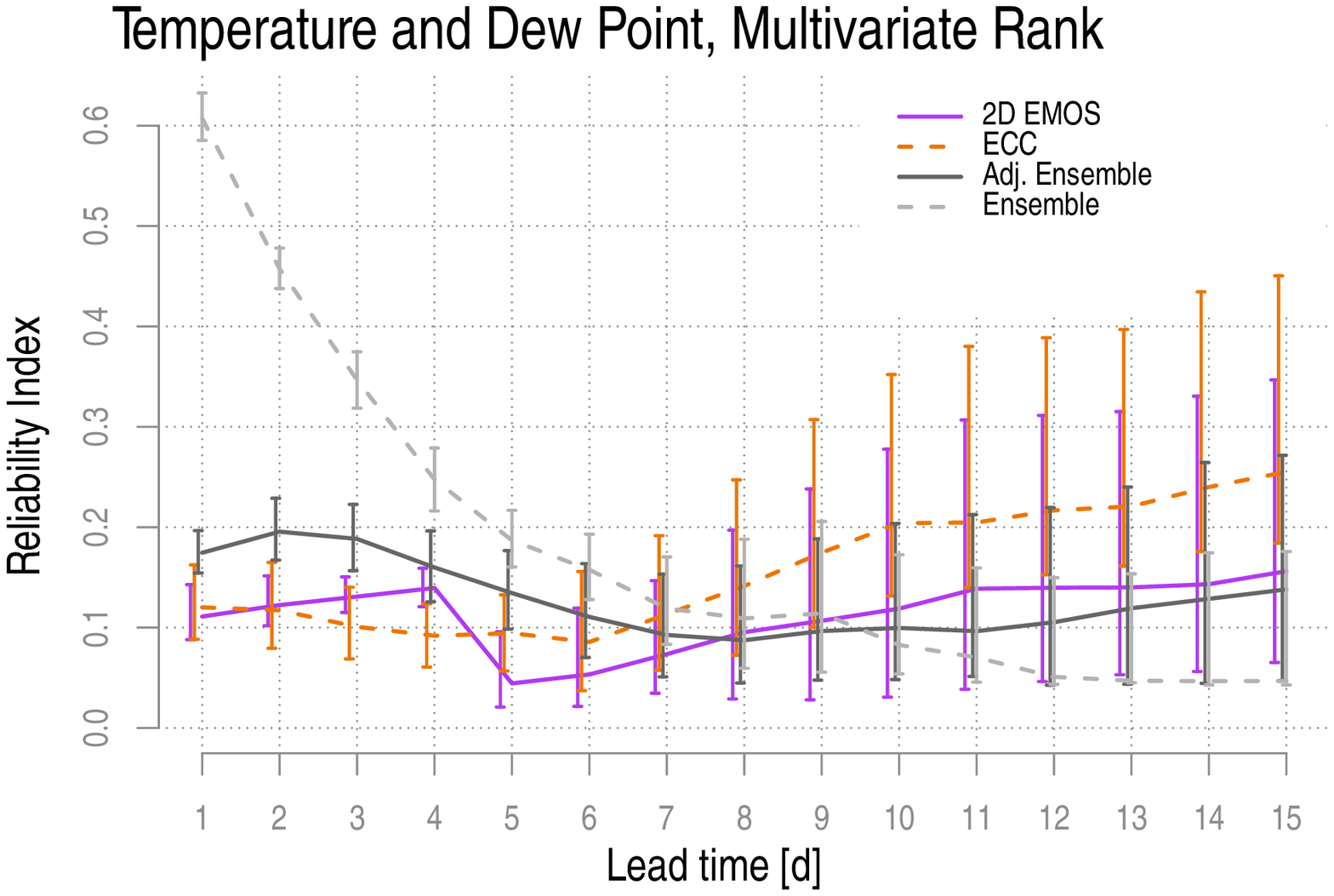, width=.47\textwidth} 
\end{center}
\caption{Reliability indices of bivariate forecasts corresponding to average ranks ({\em left}) and multivariate ordering ({\em right}) together with $95\,\%$ confidence intervals.}
\label{fig:relInd2D}
\end{figure} 

Reliability of the bivariate forecast is now investigated. Figure \ref{fig:relInd2D} presents reliability indices (corresponding to average ranking and multivariate ordering) as a function of the forecast lead time. The reliability indices are rather consistent with one another: similar patterns are found for the two types of ranking (average and multivariate).  One exception is  the difference between raw and adjusted ensemble results which is much smaller when average ranks are considered. The raw ensemble results in strongly U-shaped histograms at short lead time, while the ensemble adjustment allows to improve reliability in particular with respect to the multivariate ranking. 2D EMOS exhibits the better reliability (lowest reliability indices) up to day 8 and fairly good results for longer lead times. However, EMOS post-processed forecasts are slightly biased towards the end of the forecast range, producing low multivariate ranks of the verifying observations (see Figure \ref{fig:histAV} and \ref{fig:histMV} of Appendix \ref{secA1}). ECC forecasts are highly hump shaped at short lead times, rather well calibrated at medium lead times,  and U-shaped for long lead times. Post-processing approaches applied here fail to improve reliability of the ensemble beyond day 10, which is in line with their predictive performance in terms of the ES. 

In a nutshell, bivariate post-processing can improve multivariate forecasts of temperature and dew point, and in particular reliability, in a multivariate sense, is increased. However, the post-processing methods applied here do not explicitly account for any relationship between temperature and dew point respecting the laws of physics.  This limitation results sometimes in situations where a dew point forecast is greater than a temperature forecast. In that case, the former is set equal to the latter in order to avoid  physical inconsistencies and allow the computation of the heat indices. This situation occurs less than 3\,\% of the cases when applying 2D EMOS and can be more frequent  when simply adjusting the ensemble forecasts.

\subsection{Post-processing of discomfort index and indoor wet-bulb globe temperature forecasts}
  \label{calibDIWBGT}

Calibrated DI and WBGTid forecasts can be derived from the jointly calibrated temperature and dew point forecasts, but DI and WBGTid ensemble forecasts can also be  directly post-processed using the GEV EMOS model introduced in Section \ref{GEVEMOSmodel}. The GEV model has only $5$ free parameters in the case of a single group of exchangeable ensemble members. Considering a 60-day rolling training period and the same verification period 14 July -- 30 September 2017 as before, we inspect the CRPSS of various semi-local EMOS configurations with respect to local EMOS for DI and WBGTid (not shown). Based on this exploratory analysis, we propose to adapt the EMOS configuration as a function of the forecast lead time as follows: for both indices local EMOS is applied between day 1 and 4,  semi-local EMOS with 75 and 5 clusters for days 5 -- 8 and 9 -- 15, respectively.  This model is referred to as {\em GEV EMOS\/} (see Table \ref{tab:tab1}).

\begin{figure}[t]
\begin{center}
\epsfig{file=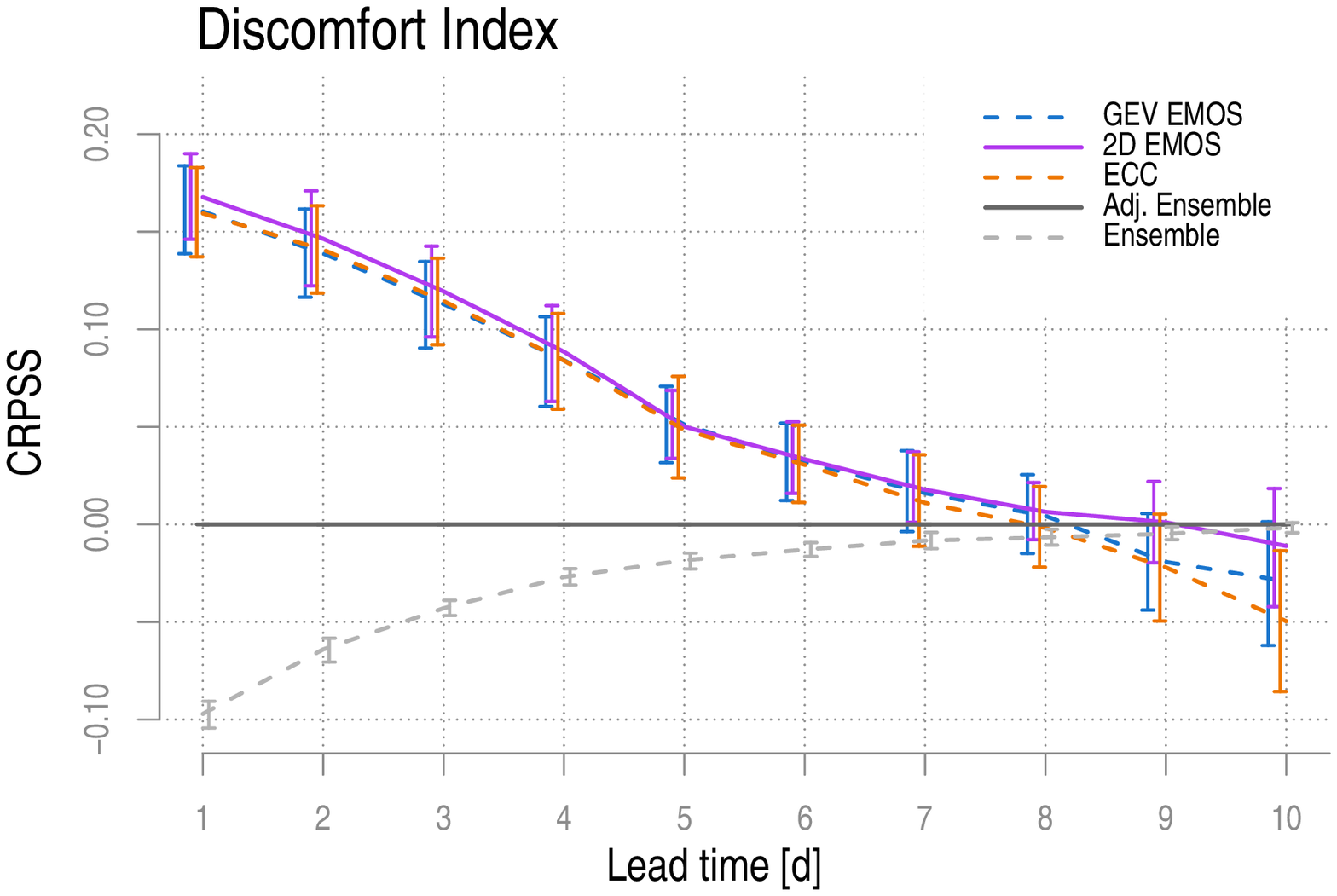, width=.47\textwidth} \qquad
\epsfig{file=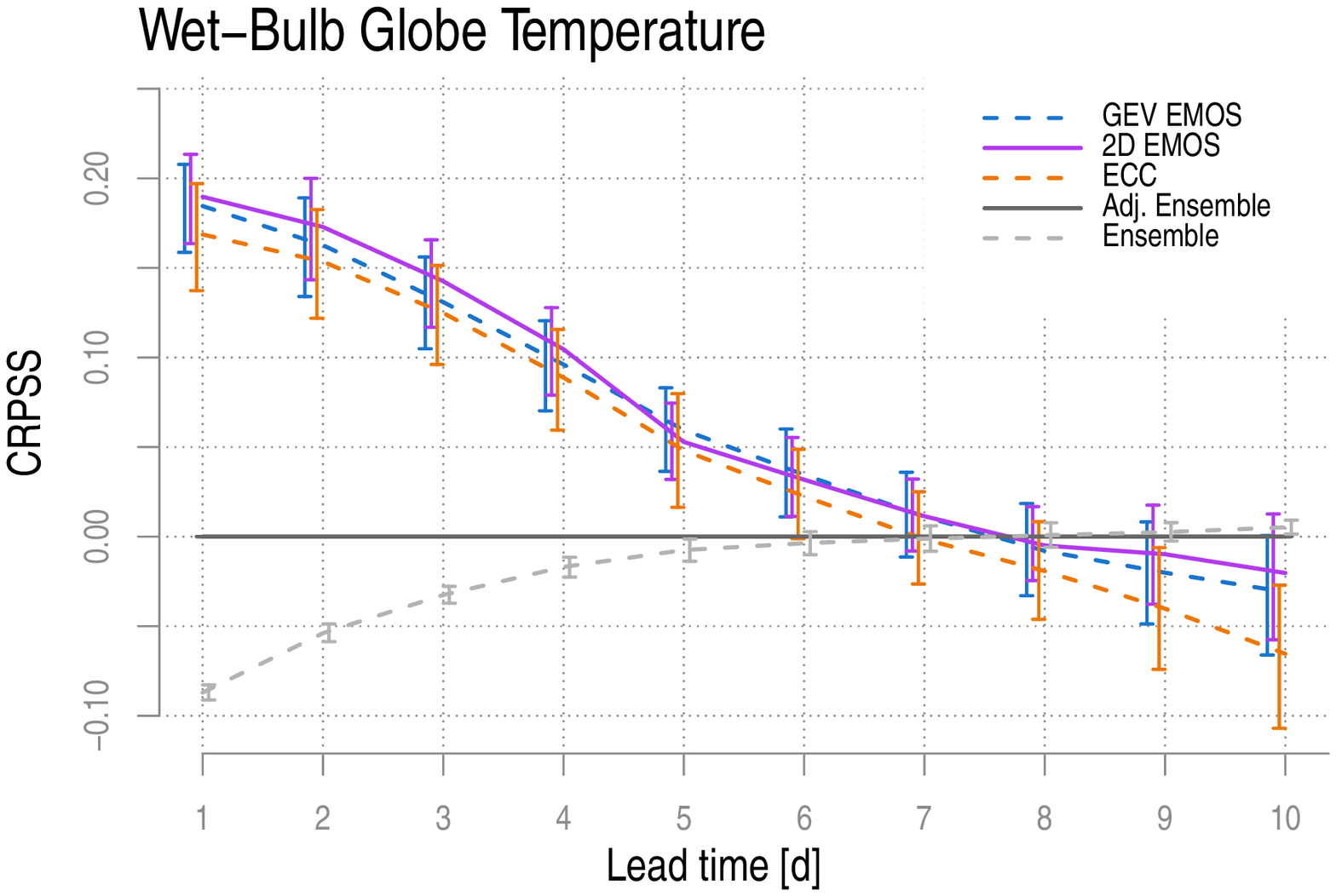, width=.47\textwidth} 
\end{center}
\caption{CRPSS with respect to the adjusted ensemble of DI ({\em left}) and WBGTid ({\em right}) forecasts together with $95\,\%$ confidence intervals.}
\label{fig:crpss}
\end{figure}  

In Figure \ref{fig:crpss}, results in terms of CRPSS  indicate that post-processing significantly improves both the raw and adjusted ensembles up to day 7 for DI and day 6 for WBGTid. For both indices, 2D EMOS post-processed forecasts  have the best predictive skill, followed closely by GEV EMOS forecasts, whereas ECC shows the worst predictive performance among the calibration approaches. However, the differences between models are minor and often non-significant. For lead times up to day 7,  the differences in CRPS between the competing EMOS forecasts are significant at a $5\,\%$ level for less than $6\,\%$ of the stations. This later number increases to more than 15\% when differences are computed against the adjusted ensemble forecasts (not shown).

Verification rank and PIT histograms help diagnosing reliability issue in the forecast. Results for DI are shown in Figures \ref{fig:histDI} of Appendix \ref{secA2}, while histograms for WBGTid, which are very similar, are not shown. For both indices, the raw ensemble forecasts are highly underdispersive for day 1; however, the dispersion improves with the forecast lead time. Moreover, raw forecasts exhibit a bimodal character, which is in line with the bivariate histograms of raw temperature and dew point forecasts. Only local GEV EMOS calibration removes this bimodality and in general, similar to bivariate forecasts, the longer the lead time, the smaller the effect of post-processing. 

\begin{figure}[t!]
\begin{center}
\epsfig{file=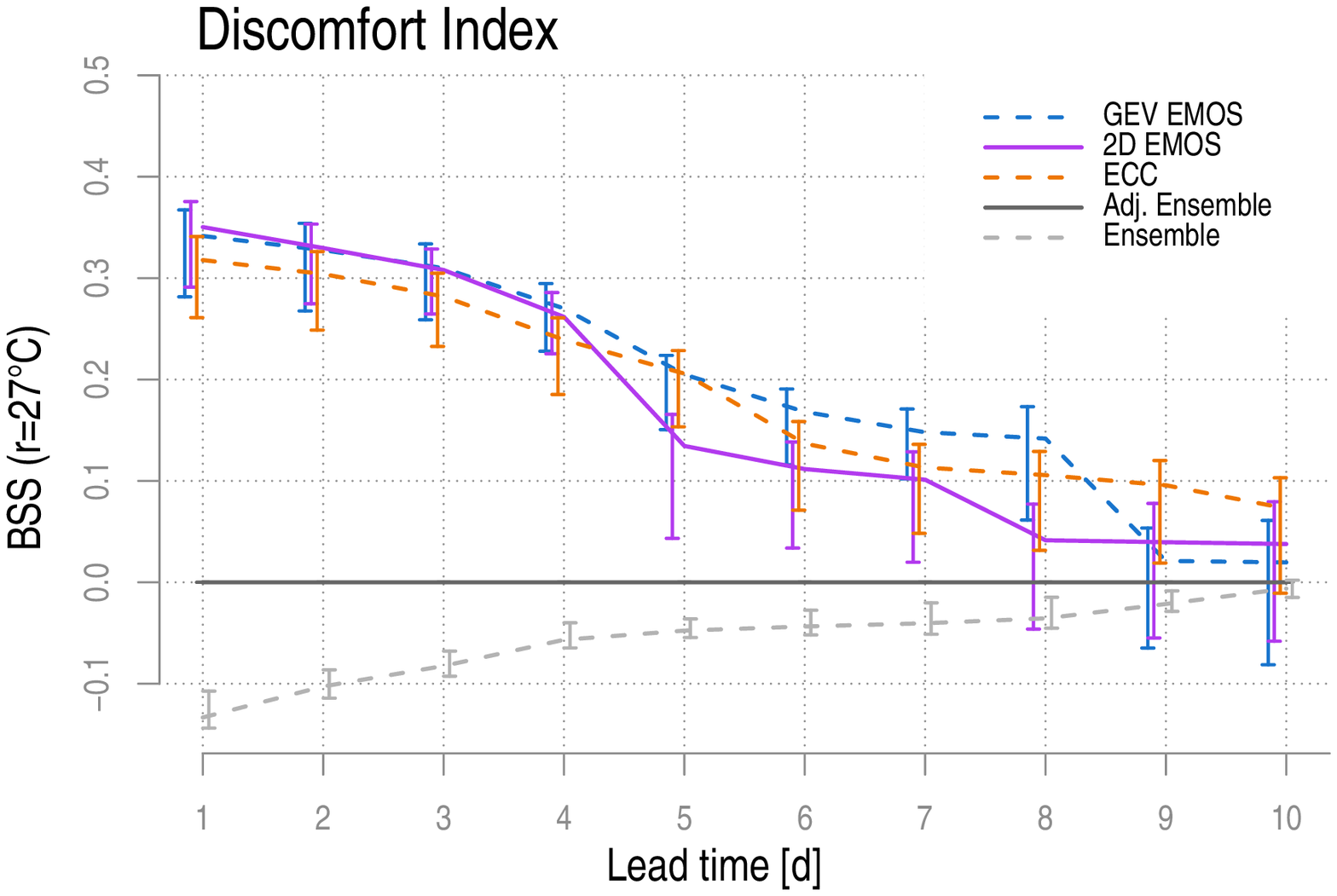, width=.49\textwidth} \
\epsfig{file=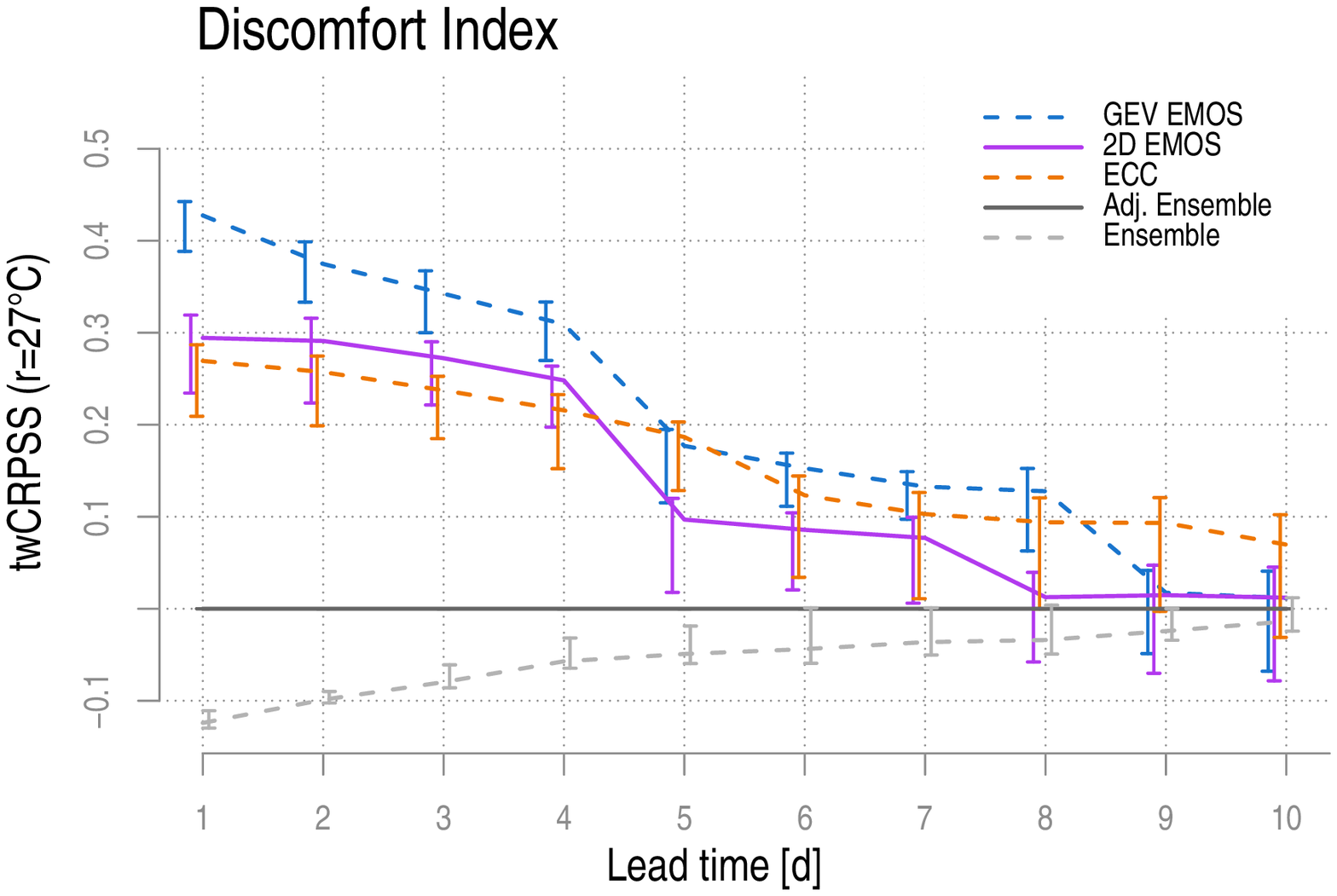, width=.49\textwidth} 
 
\smallskip
\epsfig{file=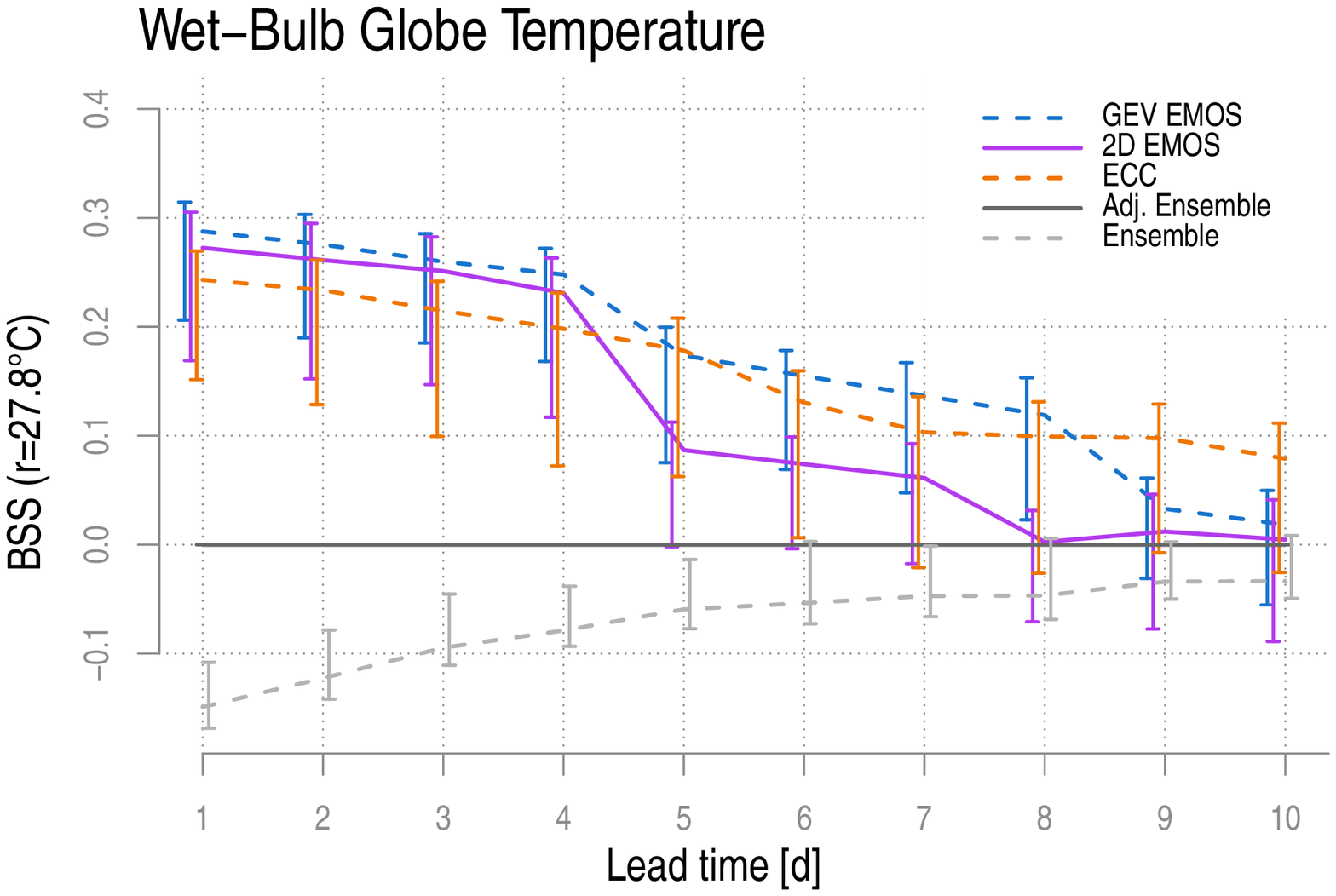, width=.49\textwidth} \
\epsfig{file=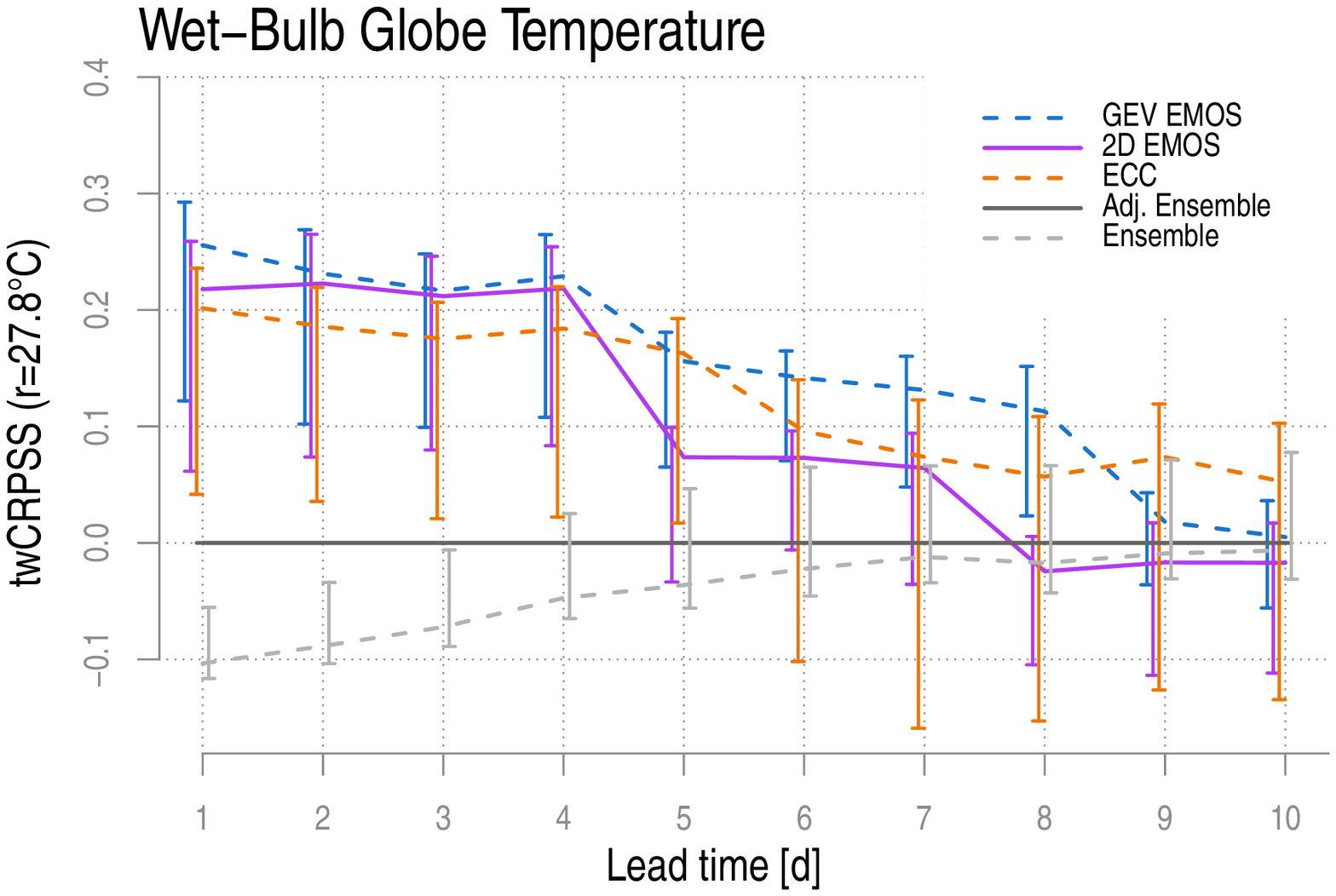, width=.49\textwidth} 
\end{center}
\caption{BSS ({\em left}) and twCRPSS ({\em right}) with respect to the adjusted ensemble corresponding to thresholds \ $r= 27\,^\circ$C \ for the DI ({\em top panel}) and  \ $r= 27.8\,^\circ$C \ for the WBGTid ({\em bottom panel}) forecasts together with $95\,\%$ confidence intervals.}
\label{fig:bsstwcrpss}
\end{figure}

Another perspective is now taken focusing on the forecast skill at the right tail of the observation distributions.  Figure \ref{fig:bsstwcrpss} (top panel) depicts BSS and twCRPSS considering the event DI exceeding \ $r= 27\,^\circ$C. \ This threshold corresponds to the level at which most of the population suffers from heat-induced discomfort \citep{scks05}.  In terms of BSS, all EMOS approaches outperform the raw and adjusted ensemble up to  day 10 with comparable skill. However, in terms of twCRPS, the GEV EMOS model clearly outperforms the other post-processing methods. For lower thresholds, the differences between the EMOS approaches is less pronounced, while for higher thresholds the results are noisier and the significance bars are larger (not shown).

For WBGTid, a threshold  \ $r= 27.8\,^\circ$C \ is considered. It corresponds to the threshold associated with the first warning level (green flag) \citep{pms13}. Figure \ref{fig:bsstwcrpss} (bottom panel) shows BSS and twCRPSS: post-processed forecasts significantly outperform the raw and adjusted ensemble forecasts up to at least day 8, whereas ECC exhibits positive skill score up to day 10. Considering larger thresholds, small BS and twCRPS values combined with large uncertainty in the scores leads to quite noisy results which are difficult to interpret (not shown). 
The change in EMOS configuration (from local to semi-local) is clearly visible in Figure \ref{fig:bsstwcrpss}: jumps appear at day 5 and 8 for 2D EMOS,  at day 5 and 9 for GEV EMOS, while the curves corresponding to ECC remain smooth. 

\begin{figure}[t!]
\begin{center}
\epsfig{file=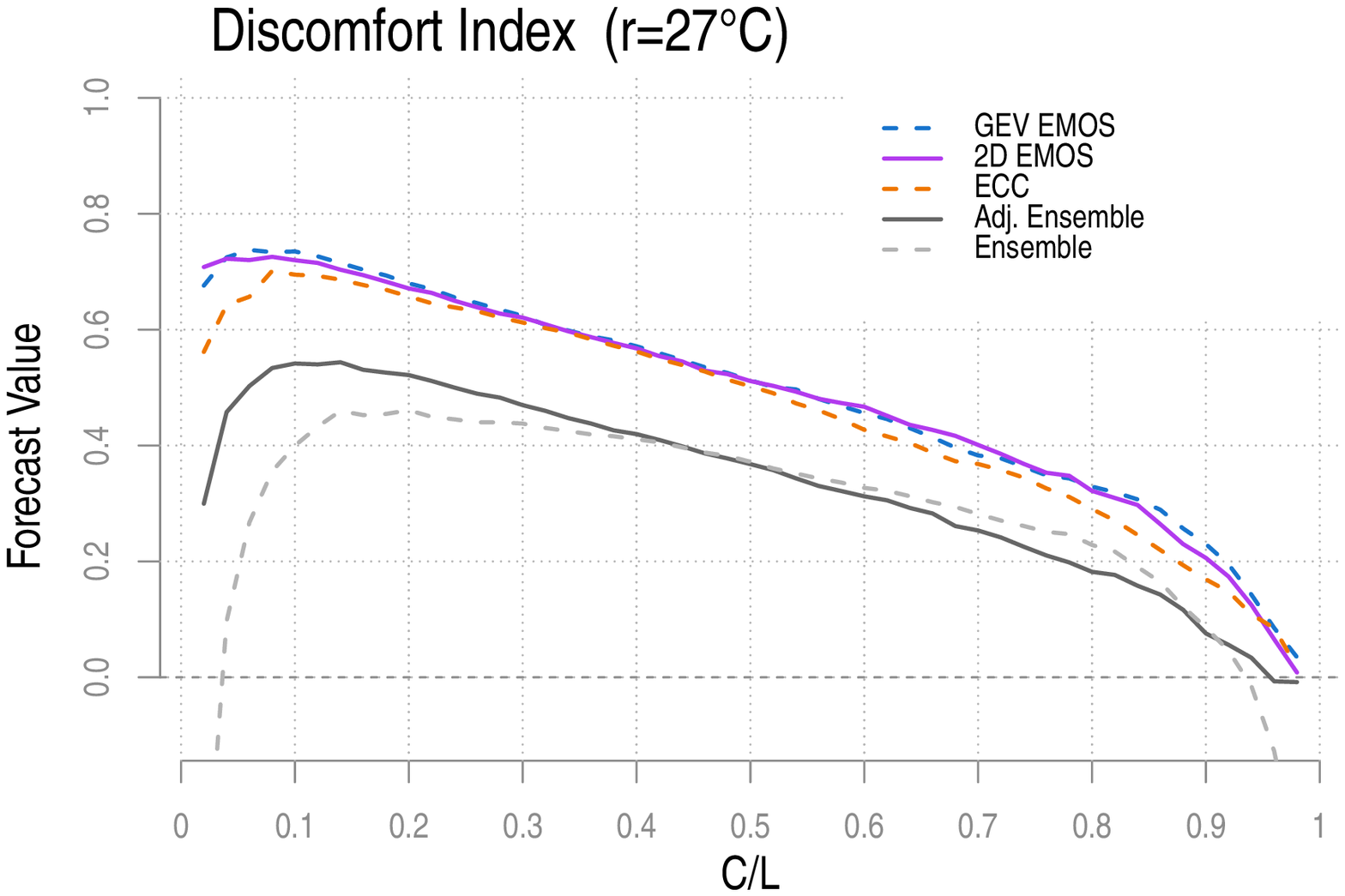, width=.49\textwidth} \
\epsfig{file=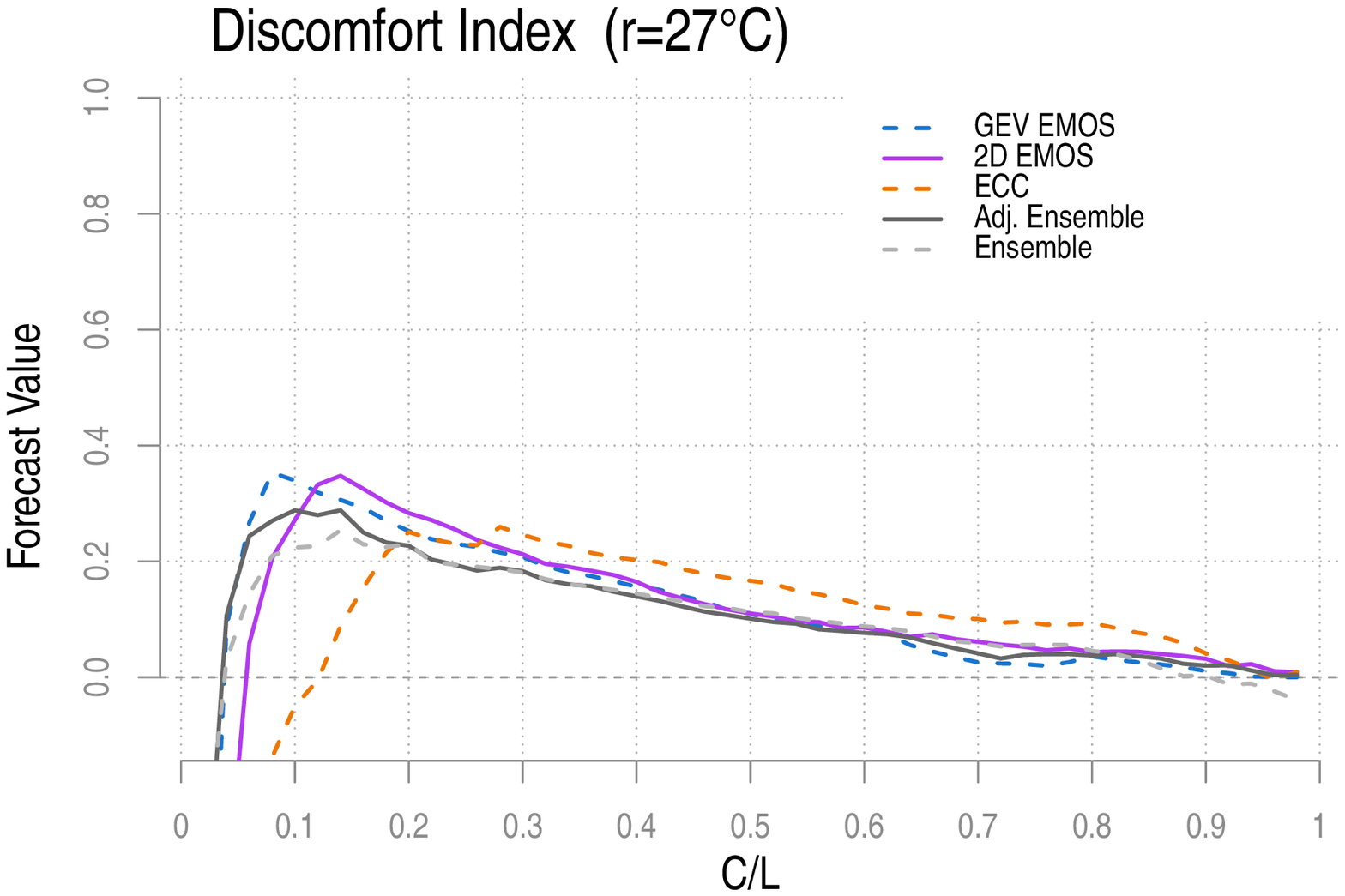, width=.49\textwidth} 

\smallskip
\epsfig{file=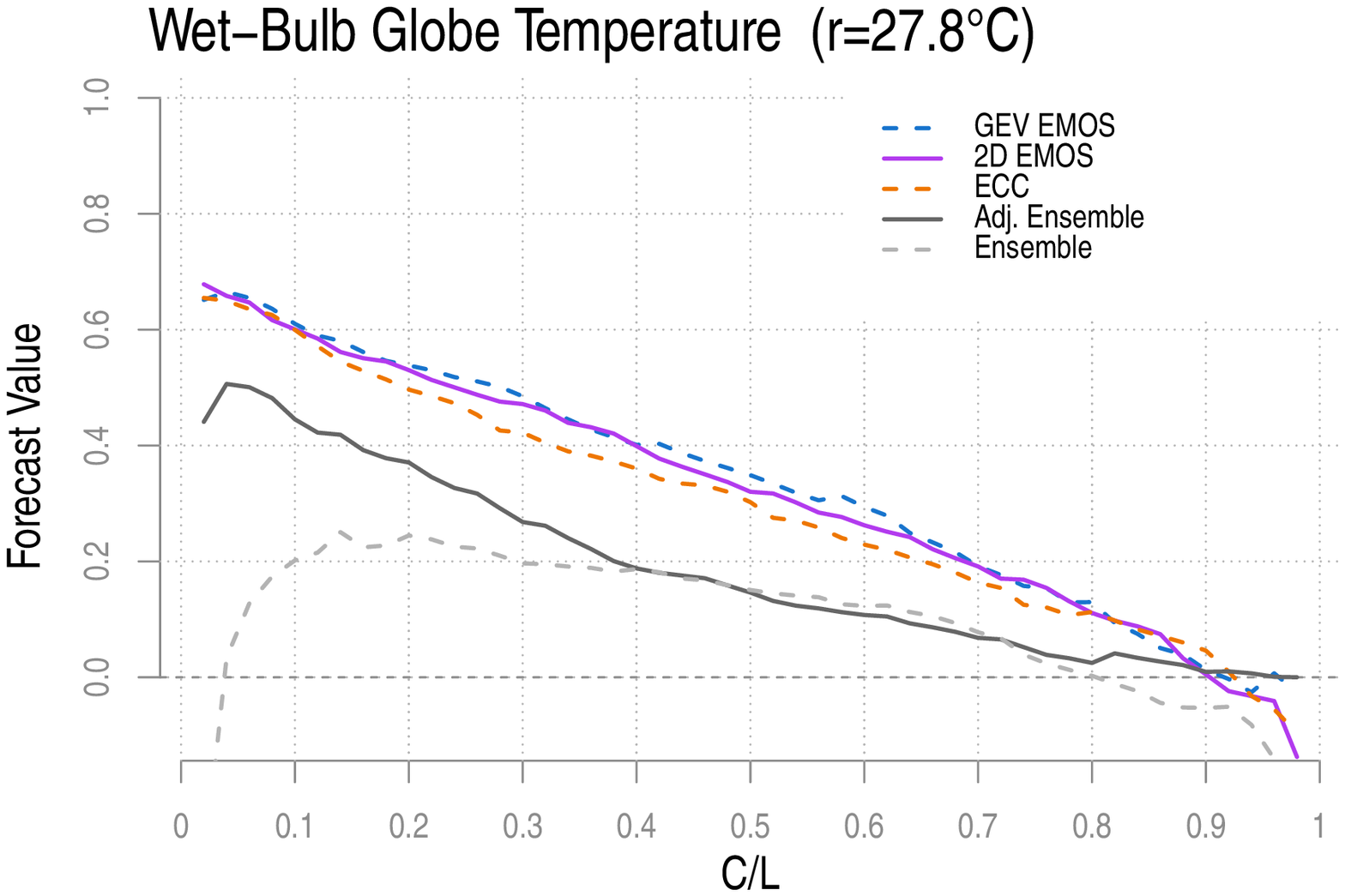, width=.49\textwidth} \
\epsfig{file=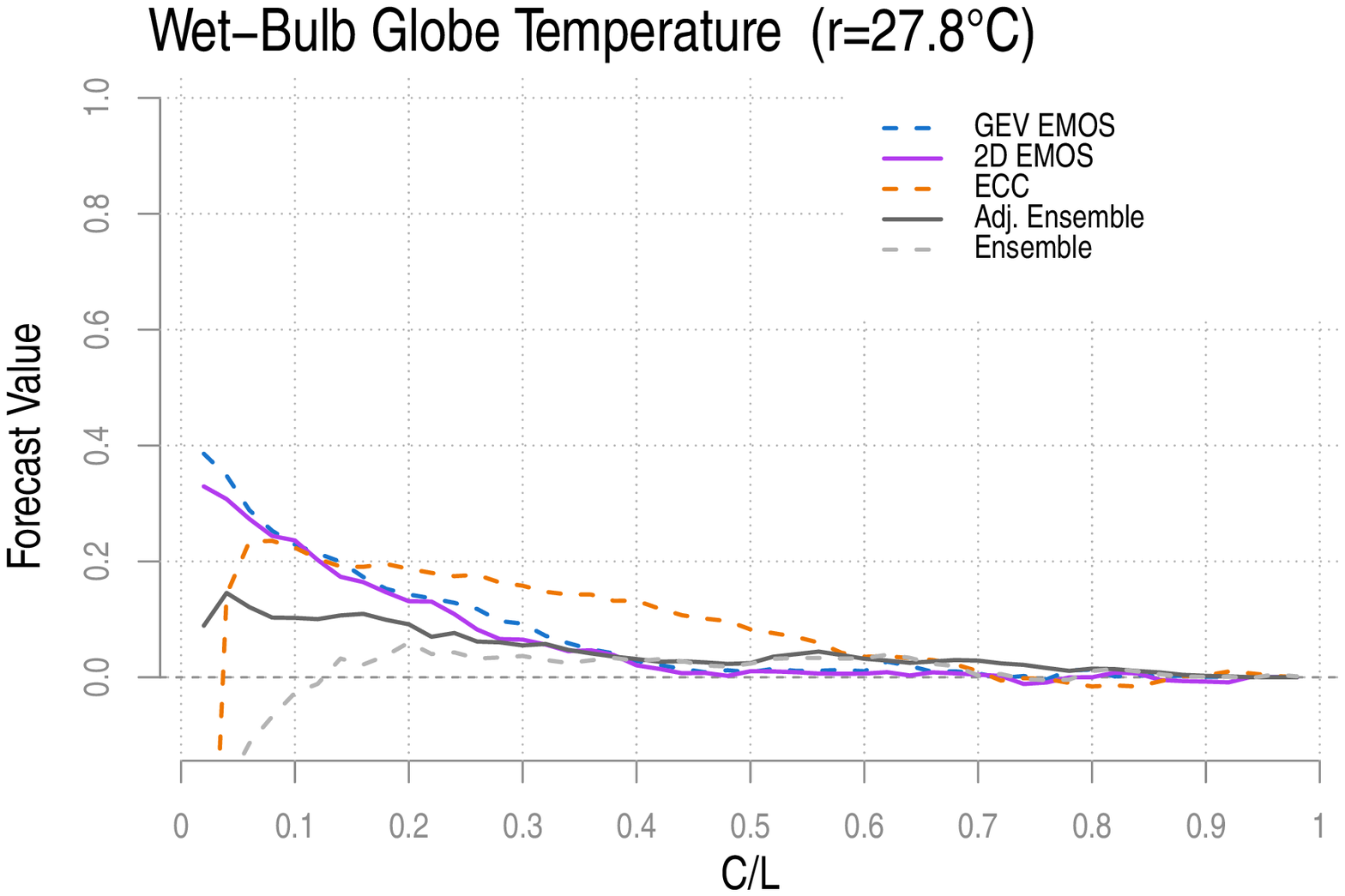, width=.49\textwidth} 
\end{center}
\caption{Aggregated forecast value for day 4 ({\em left}) and day 10 ({\em right})
corresponding to thresholds \ $r= 27\,^\circ$C \ for the DI ({\em top panel}) and  \ $r= 27.8\,^\circ$C \ for the WBGTid ({\em bottom panel}) forecasts.}
\label{fig:fv}
\end{figure}

One more step towards refined performance analysis is taken by looking at the forecast value. For the two events considered above, forecast value is plotted as a function of the user cost-loss ratio in Figure \ref{fig:fv}. As a general feature, not only the forecast value, but also the number of users that can benefit from the forecast, decreases with the forecast lead time. At day 10, the forecast of WBGTid has no value for a user who needs high certainty about the event occurrence (high cost-loss ratios) but it is still valuable for a user who is risk adverse (low cost-loss ratios). The post-processed forecasts have value for some users even beyond day 10 (not shown).  As a purely statistical characteristic of the score, the forecast value reaches a maximum for a user with cost-loss ratio equal to the climatological base rate. This is true locally, but the maximum value varies from one location to another and the corresponding pick is therefore smoothed out when aggregated over different locations. In any case, the event for DI is more probable than the event for WBGTid, which explains that the maximum is reached for different \ $C/L$ \ values in the two cases. The raw ensemble has negative value for extreme \ $C/L$ \ for both DI and WBGTid forecasts at 4 and 10 days, but the ensemble adjustment allows to make the forecasts valuable for (almost) all types of users. Forecast post-processing increases the forecast value further in most of the cases. There are also slight differences in terms of forecast value between the different calibration approaches.

\subsection{Predictive distributions of categories of heat indices}
  \label{categories}
  
\begin{table}[t]
  \begin{center}{\footnotesize
  \begin{tabular}{c|l|l|c|l|l}
  \multicolumn{3}{l|}{Discomfort Index ($I_d$)}&\multicolumn{3}{l}{Wet-Bulb Globe Temperature ($T_{WBGid}$)}\\ \hline
  Cat.& Range ($^\circ$C)&Classification&Cat.& Range ($^\circ$C)&Flag col.\\ \hline
 1& $I_d \!<\!21$& No discomfort&1 & $T_{WBGid}\!<\! 27.8$& No flag   \\
 2& $21\!\leq\! I_d \!<\!24$& Under $50\,\%$ population feels discomfort&2 & $27.8 \!\leq\! T_{WBGid}\!<\! 29.5$& Green \\
 3& $24\!\leq\! I_d \!<\!27$& Over $50\,\%$ population feels discomfort&3 & $29.5 \!\leq\! T_{WBGid}\!<\! 31.1$& Yellow \\
 4& $27\!\leq\! I_d \!<\!29$& Most of population feels discomfort&4 & $31.1 \!\leq\! T_{WBGid}\!<\! 32.2$& Red  \\
 5& $29\!\leq\! I_d \!<\!32$& Everyone feels severe stress&5 & $T_{WBGid}\!\geq\! 32.2$& Black \\
 6& $I_d \!\geq\! 32$&State of medical emergency&&& \\\hline 
  \end{tabular}}
\end{center}
\caption{Classification of DI and heat categories of WBGTid based on \citet{scks05} and \citet{pms13}, respectively. \label{tab:tab2}}
\end{table}  

Heat indices are often grouped into categories according to the severity of the effect of heat stress on human activity. Table \ref{tab:tab2} provides a classification of DI and heat categories of WBGTid based on \citet{scks05} and \citet{pms13}, respectively. Instead of continuous heat indices, we focus now on the corresponding categories taking values \ $\{1,2,3,4,5,6\}$ \ for DI and \ $\{1,2,3,4,5\}$ \ for WBGTid. In this case, a predictive distribution is specified by a probability mass function (PMF) and post-processing reduces to a classification problem, which is a typical application area of machine learning methods. Here, predictive distributions are obtained with the help of multilayer perceptron neural networks \citep[MLP NN;][]{htf09}. We skip the intermediate step of obtaining DI and WBGTid ensemble forecasts from the corresponding forecasts of temperature and dew point and connect the latter directly to the DI and WBGTid categories. As primary input variables (also called features or covariates), we use the mean, variance and covariance of temperature and dew point ensemble forecasts, as well as the geographical coordinates (latitude, longitude), and the elevations of the SYNOP stations and of the corresponding model grid point. Instead of applying local or semi-local training, all forecasts cases in the training periods are regarded as a single group.  This nine-dimensional feature vector can be extended further with forecast errors of temperature and dew point.  Ensemble mean errors of the 1 to 5 closest days to the end of the training period can be added to the input variables increasing the number of covariates by 2 to 10 more features. 
In this case, the length of the training period is naturally reduced by the number of days where the forecasts errors are considered. 

\begin{table}[t]
\begin{center}{\footnotesize
\begin{tabular}{l|l|c|c|c|c|c|c|c|c|c|c|c|c|c|c|c|}
\multicolumn{2}{l|}{MLP NN}&\multicolumn{15}{c}{Lead time (day)}\\ \cline{3-17}
\multicolumn{2}{l|}{configurations}&1&2&3&4&5&6&7&8&9&10&11&12&13&14&15\\ \hline
&Net&\multicolumn{4}{c|}{single}&\multicolumn{4}{c|}{single}&\multicolumn{7}{c|}{dual} \\ \cline{2-17}
DI&Neurons in hidden layer&\multicolumn{4}{c|}{40}&\multicolumn{4}{c|}{15}&\multicolumn{7}{c|}{15--15} \\ \cline{2-17}
&Forecast errors&\multicolumn{4}{c|}{5}&\multicolumn{4}{c|}{0}&\multicolumn{7}{c|}{0} \\ \hline
&Net&\multicolumn{8}{c|}{single}&\multicolumn{4}{c|}{single}&\multicolumn{3}{c|}{single} \\ \cline{2-17}
WBGTid&Neurons in hidden layer&\multicolumn{8}{c|}{30}&\multicolumn{4}{c|}{35}&\multicolumn{3}{c|}{30} \\ \cline{2-17}
&Forecast errors&\multicolumn{8}{c|}{5}&\multicolumn{4}{c|}{0}&\multicolumn{3}{c|}{0} \\ \hline
\end{tabular}
}
\end{center}
\caption{MLP NN configurations for different lead times. \label{tab:tab3}}
\end{table}

Predictive PMFs of the DI and WBGTid categories are obtained applying MLP NNs with a single hidden layer and cross-entropy as loss function. In our case study, the different categories are highly imbalanced: the proportions of observations in the first category in Table \ref{tab:tab2} is $\sim 67\,\%$ for DI and  $\sim 99\,\%$  for WBGTid, whereas the last category in Table \ref{tab:tab2} represents less than $1\,\permille$  in both cases.  For this reason, we also consider a two step procedure (referred to as {\em dual net\/}).  In the first step a binary classification is performed, which
for each feature vector results in the distribution \ $(p_1,1-p_1)$ \ of the binary outcome of belonging to the first (richest) category or not. Then a second net is trained using only feature vectors coming from the remaining \ $\kappa$ \ classes, where \ $\kappa=5$ \ for DI and  \ $\kappa=4$ \ for WBGTid. For each feature set this step yields a distribution \ $(q_1,\ldots ,q_{\kappa})$ \ of the corresponding classes and the final distribution of index categories is obtained as \ $\big(p_1,(1-p_1)q_1, \ldots, (1-p_1)q_{\kappa}\big)$.

Similar to the EMOS approaches, different training configurations are envisaged for different lead times. We tested the performance of single nets with 15, 20, 25, 30, 35, 40 and 50 neurons in the hidden layer, both with the nine-dimensional primary features and with extended feature vectors containing forecast errors of either 1 or 5 days. In the case of dual nets, we used 15, 25 and 40 neurons in the hidden layer of both components and considered the primary features plus the 1 day forecast error. In general, a large number of neurons leads to overfitting, and according to our observations, for longer lead times the use of forecast errors does not improve the predictive performance, while the dual net can handle better the larger uncertainty in the forecasts. For DI categories, the following configuration is selected:  a single net with 40 neurons making use of the forecast errors for 5 days as additional predictors for days 1 -- 4, single net with 15 neurons for days 5 -- 8, and  dual net with 15 -- 15 neurons for days 9 -- 15 (see Table \ref{tab:tab3}). For WBGTid categories, single networks with medium number of neurons are preferred, and moreover, the use of forecast errors has a longer lasting effect. So, the selected configuration is the following: for days 1 -- 8 we consider 30 neurons and 5-day forecast errors, for days 9 -- 12 only the primary features and 35 neurons, whereas for the last 3 days we just reduce the number of neurons to 30 (see Table \ref{tab:tab3}).  In the following discussion these two blends are referred to  as {\em MLP NN\/}. 

\begin{figure}[t]
\begin{center}
\epsfig{file=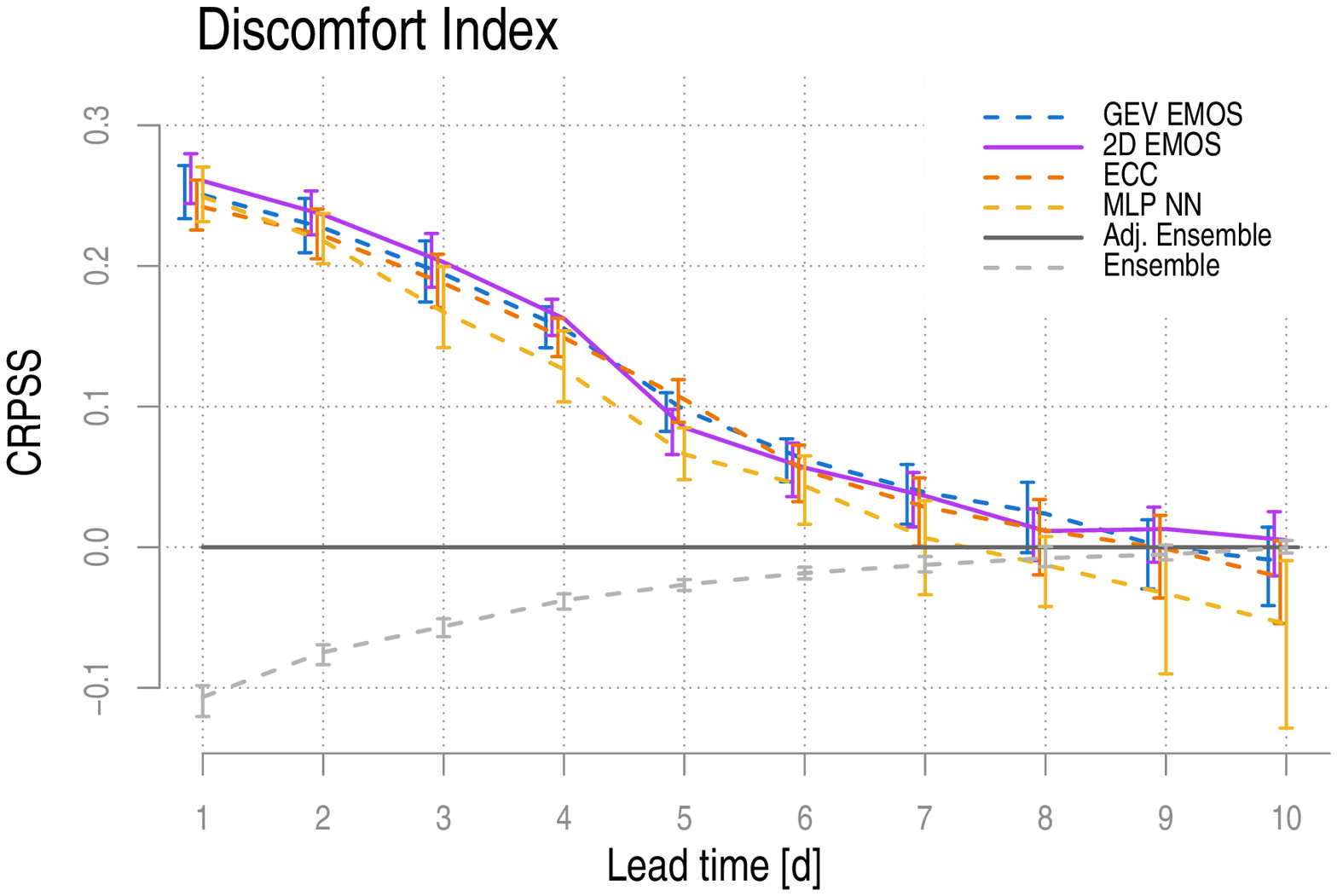, width=.47\textwidth} \qquad
\epsfig{file=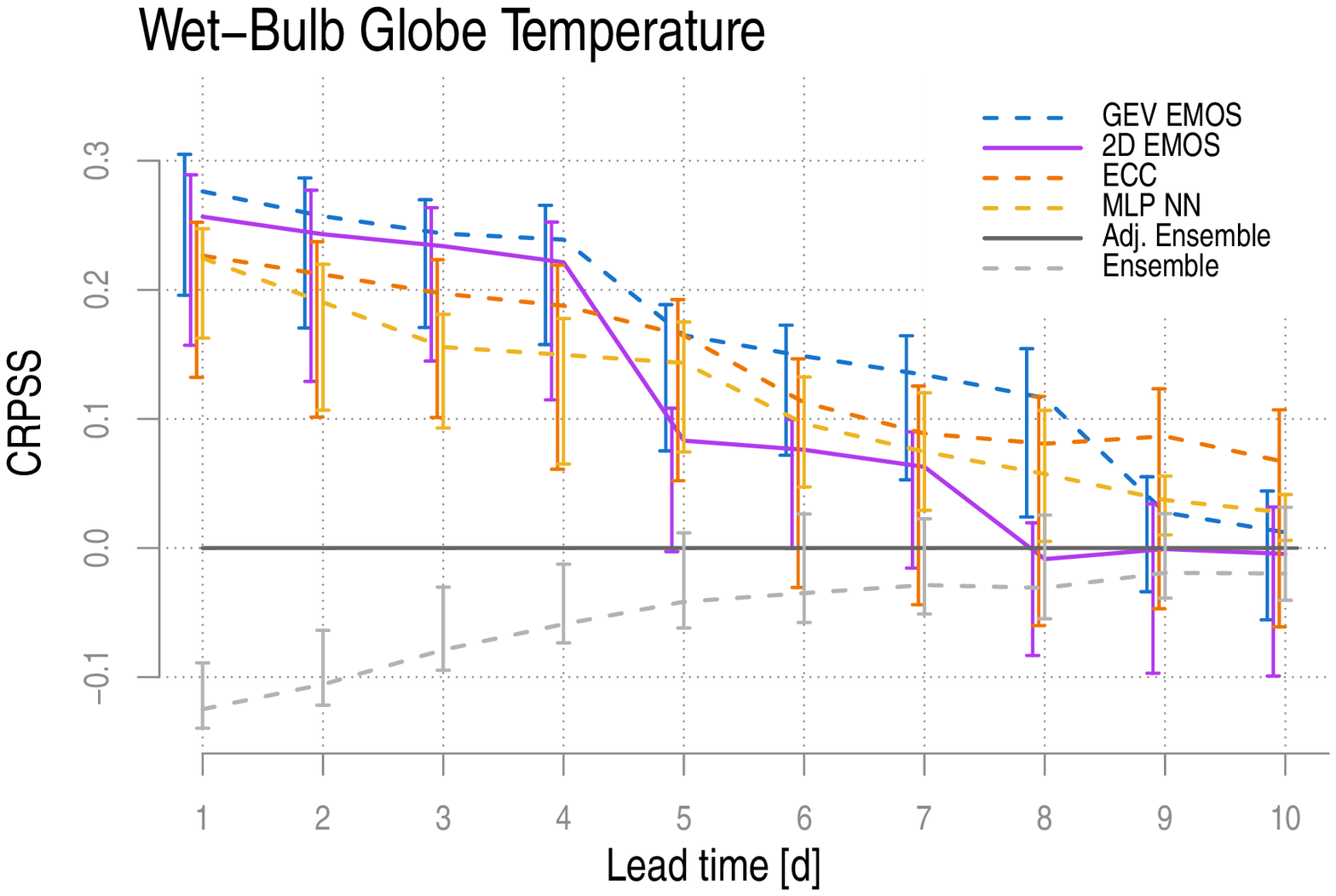, width=.47\textwidth} 
\end{center}
\caption{CRPSS with respect to the adjusted ensemble of DI ({\em left}) and WBGTid ({\em right}) categories together with $95\,\%$ confidence intervals.}
\label{fig:crpssNN}
\end{figure}  

In Figure \ref{fig:crpssNN}, CRPS is applied to forecast and observation categories. All calibration approaches significantly outperform both the raw and the adjusted forecasts of DI categories till day 6.  The results for the different EMOS models are in line with those of the continuous forecasts (Figure \ref{fig:crpss}, left panel). The robust and rather generic MLP NN method is outperformed by the parametric models; however, the differences are often non-significant. 
A different ranking of the calibration methods can be observed in the right panel of Figure \ref{fig:crpssNN}, which shows a high level of similarities with BSS and twCRPSS plots in the bottom panel of Figure \ref{fig:bsstwcrpss}. 
The GEV and 2D EMOS forecast skill jumps at the lead times where the training configurations are changed. After day 7, 2D EMOS loses its skill against the adjusted ensemble while 
the machine learning approach shows a reasonable predictive performance up to day 10.  Among the different post-processing approaches,  MLP NN provides the sharpest predictive PMF in terms of the average width of the $90\,\%$ central predictive intervals and in terms of variance for longer lead times too (not shown). Due to the extremely imbalanced nature of WBGTid observation categories, during the training of a neural net one can easily face the situation where some categories are completely missing from the training data. This problem is less pronounced for the other investigated post-processing methods as the EMOS predictive PMFs are calculated using continuous forecasts and observations. Thus, the WBGTid results of the different classification approaches should be interpreted cautiously.

\section{Conclusion}
\label{conclusion}

This study focuses on improving forecast heat indices such as discomfort index (DI) and indoor wet-bulb globe temperature (WBGTid) driven by temperature and dew point forecasts as inputs through statistical calibration. Post-processing based on ensemble model output statistics (EMOS) approaches enable us to significantly improve continuous forecasts of heat indices, at least up to day 6. In particular, DI and WBGTid forecast reliability is clearly increased in that case. Among different flavours of EMOS approaches, the generalized extreme value (GEV) model has the best overall performance, followed by the bivariate EMOS and ensemble copula coupling. The GEV EMOS is tailored to the end products (DI and WBGTid), whereas calibration based on joint post-processing of temperature and dew point forecasts can be applied to any heat index depending on these two weather quantities.
For longer forecast ranges (typically beyond the first week in the forecast), post-processing does not improve the forecast and can even lead to a deterioration of the forecast skill which is explained by the non-stationarity of the forecast errors. 

Moving from continuous forecasts to predicted categories of heat indices, we further compare the traditional approach of post-processing to multilayer perceptron neural networks, demonstrating that both methods have comparable predictive power. Furthermore, we evidenced that  local EMOS post-processing is outperformed by cluster-based EMOS approaches after day 4 if a limited training period is used. Decreasing the number of clusters with the forecast lead time helps extending the benefit of EMOS post-processing for longer lead times. As a limit, a single cluster approach is equivalent to a global training approach which is not suitable for forecasting over a large domain.

The present work highlights  several avenues of potential future research. For example, as an alternative training data set to the one above, one could consider the use of reforecasts. This is not investigated here but should be considered to address efficiently the calibration of medium to extended range forecasts.  When accessing large training data sets, appropriate predictors could be defined to account for seasonality and/or weather regime dependent forecast errors. In this situation, machine learning approaches seem more flexible to deal with an increased number of predictors. 

The assessment of forecast performance spans from the multivariate ES to the forecast value for a range of users defined by their cost-loss ratio, from the generic univariate CRPS to a focus on extreme events with BS and twCRPS. The cascade of verification results indicate that the  ranking of the post-processing methods might differ from one score to another: the best method for one aspect of the forecast performance is not necessary the best solution for all applications. While an improvement with EMOS post-processing is expected to be effective up to day 6, the ensemble demonstrates potential skill beyond day 10 in terms of forecast value for some users. 

The comparison of bivariate calibration approaches with direct post-processing leads us to formulate the following recommendations. Direct post-processing is generally more efficient; targeting one specific event/user could be even more appropriate. However, post-processing of the input variables appears to be a competitive approach. In that case, calibration of the dependence structure gives the best results. The simplest approach, which consist in the calibration of each component separately combined using  the empirical copula of the ensemble, has reasonable performance as well.  For more complex heat indices,  calibration of the input variables can be envisaged if no EMOS model exists for the end product, but as the above discussion tends to indicate, there is no royal road to post-processing of heat indices.

\bigskip
\noindent
{\bf Acknowledgments.} \ S\'andor Baran was supported by the National Research, Development and Innovation Office under Grant No. NN125679. \'Agnes Baran and 
S\'andor Baran also acknowledge the support of the EFOP-3.6.2-16-2017-00015 project. The project was co-financed by the Hungarian Government and the European Social Fund. The authors are indebted to Claudia Di Napoli and David S. Richardson for their valuable suggestions and remarks.

\newpage
\appendix
\section{Rank histograms of bivariate models}
  \label{secA1}
  
\begin{figure}[ht!]
\begin{center}
\epsfig{file=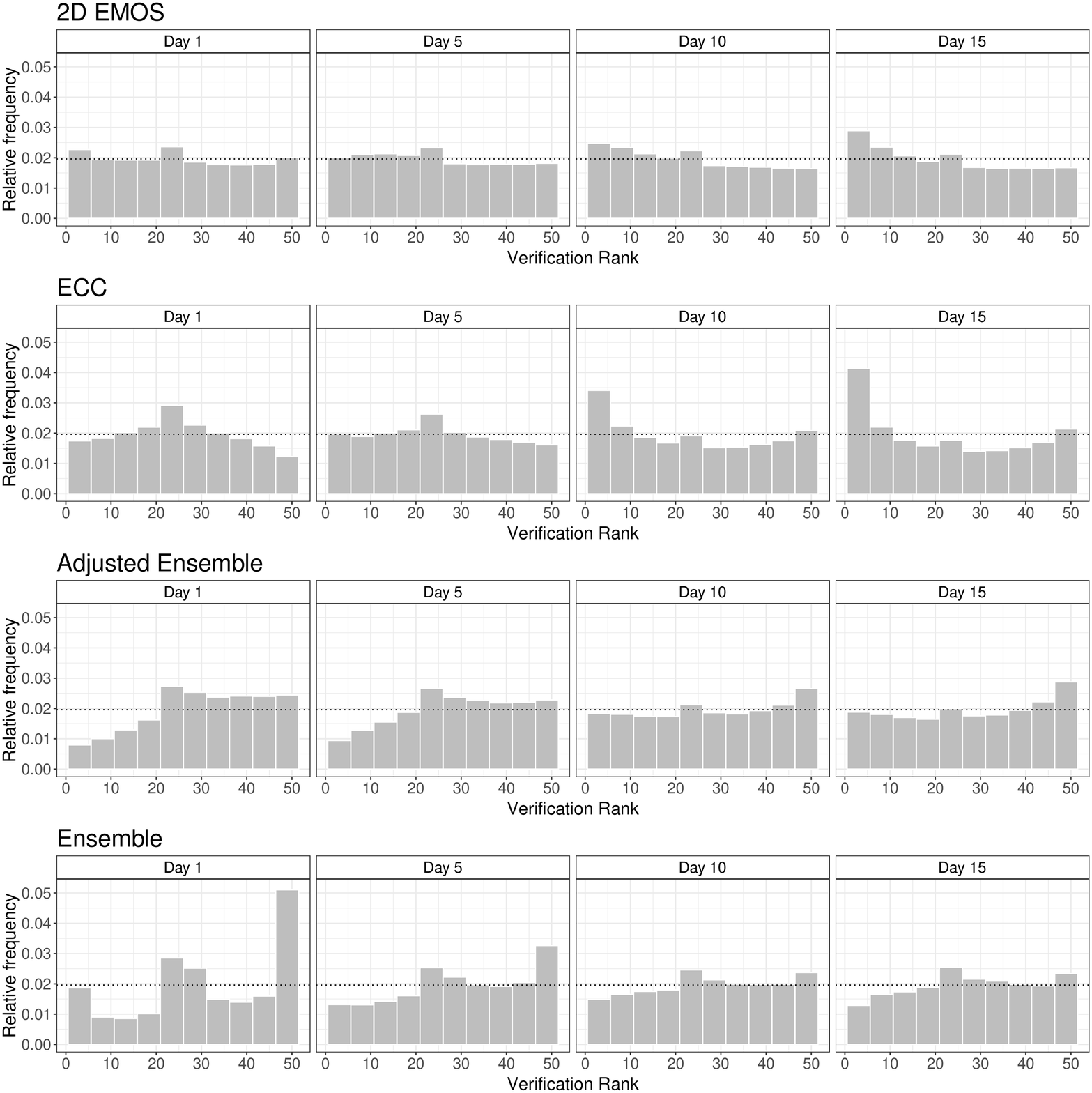, width=.99\textwidth}
\end{center}
\caption{Rank histograms of bivariate forecasts with respect to average ranks for days 1, 5, 10 and 15.}
\label{fig:histAV}
\end{figure}  

\begin{figure}[ht!]
\begin{center}
\epsfig{file=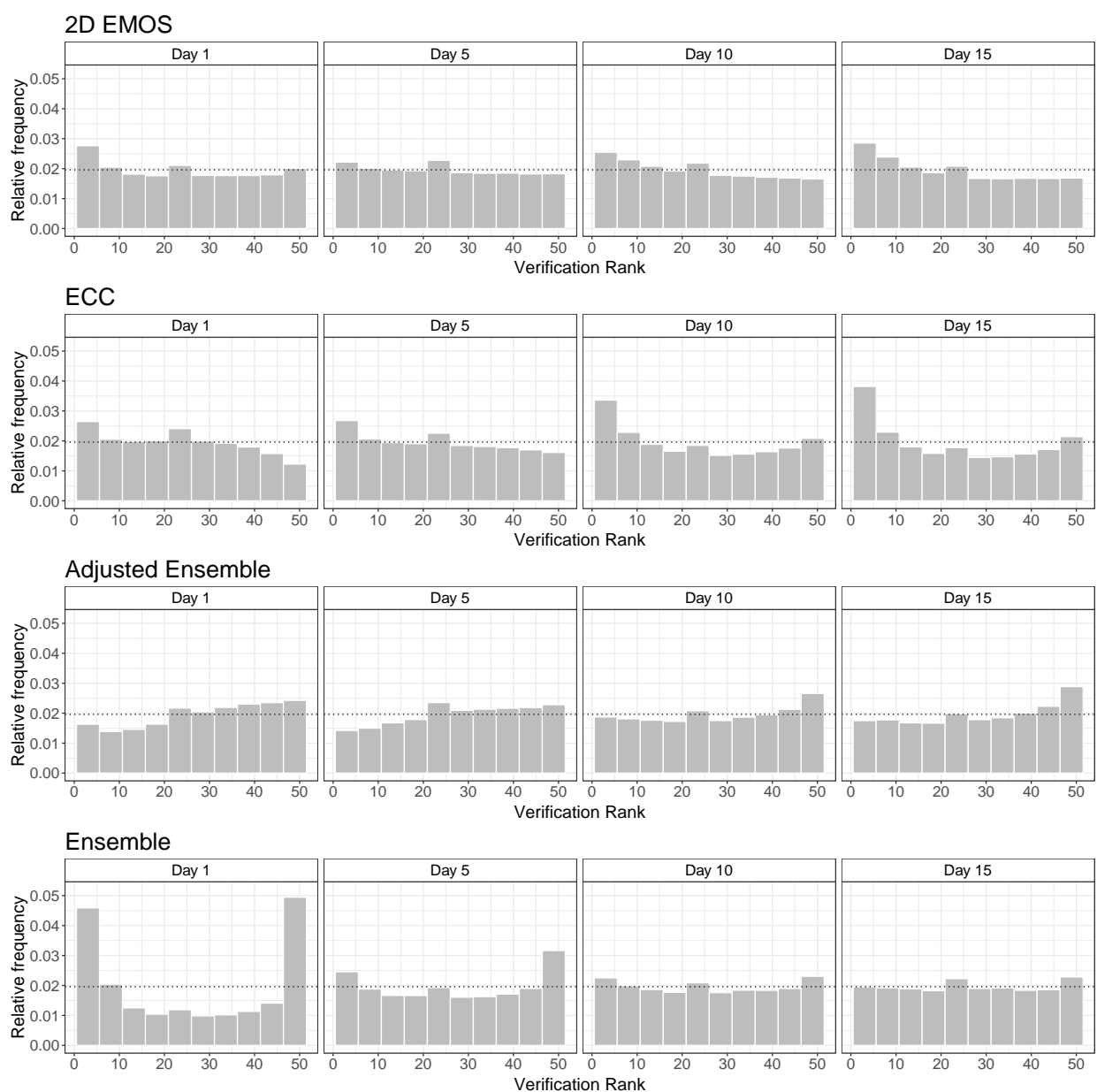, width=.99\textwidth}
\end{center}
\caption{Rank histograms of bivariate forecasts with respect to multivariate ordering for days 1, 5, 10 and 15.}
\label{fig:histMV}
\end{figure}

\newpage 
\section{Rank and PIT histograms of univariate models}
  \label{secA2}
  
\begin{figure}[ht!]
\begin{center}
\epsfig{file=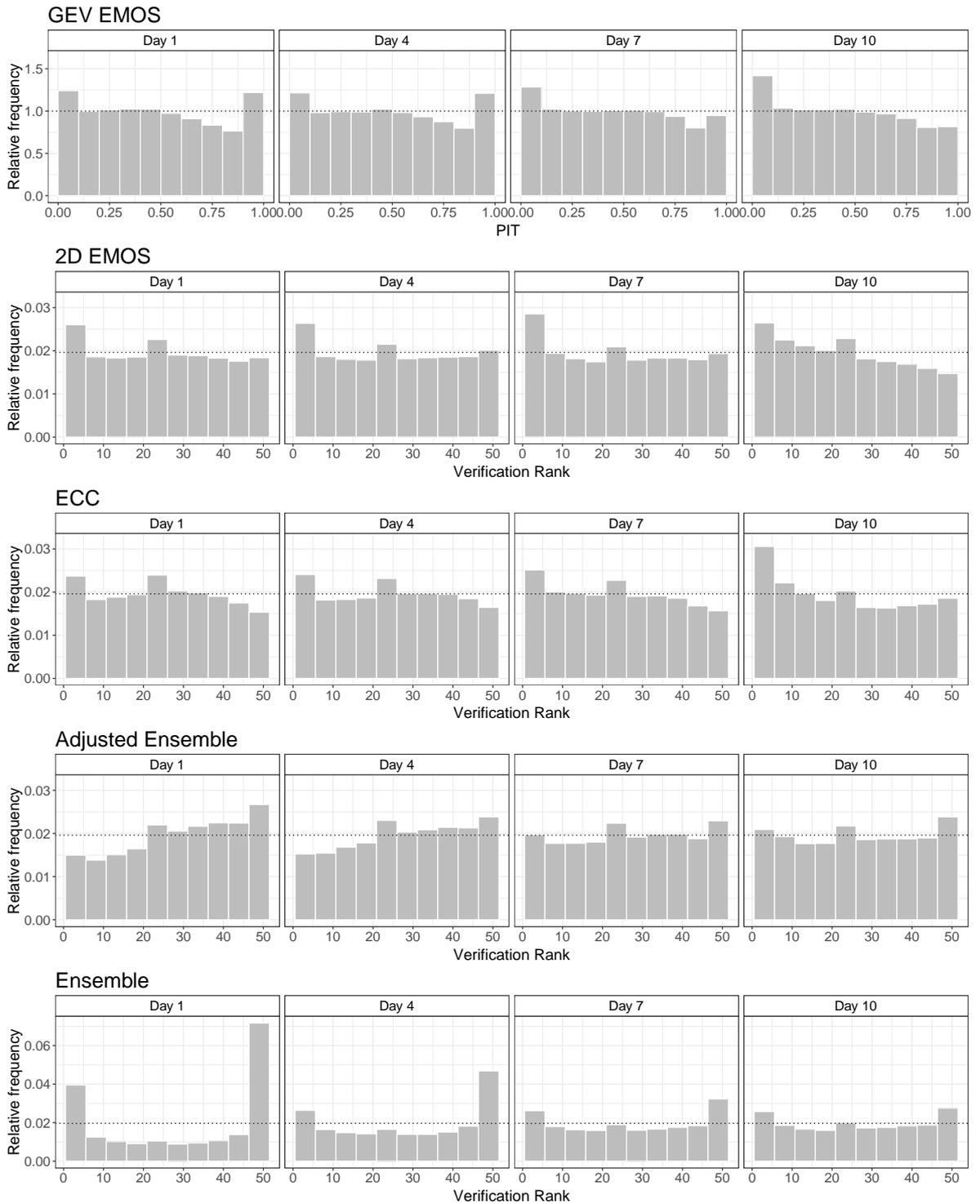, width=.95\textwidth}
\end{center}
\caption{PIT histograms of the GEV EMOS post-processed, simulated verification rank histograms of the 2D EMOS post-processed, and verification rank histograms of the adjusted and raw DI ensemble forecasts for days 1, 4, 7 and 10.  }
\label{fig:histDI}
\end{figure}  

\end{document}